\documentclass{aa}  
\usepackage{graphicx}
\def\arcsec{$^{\prime\prime}$}
\def\arcmin{$^{\prime}$}
\def\degrees{$^{\circ}$}

\def\absrmm{$\arrowvert \langle {\rm RM} \rangle  \arrowvert$}

\begin{document}
   \title{Rotation Measures of Radio Sources in Hot Galaxy Clusters}
   \subtitle{}

   \author{F. Govoni\inst{1}
          \and
          K. Dolag\inst{2}
          \and
          M. Murgia\inst{1}
          \and
          L. Feretti\inst{3}
          \and
          S. Schindler\inst{4}
          \and
          G. Giovannini\inst{3,5}
          \and
          W. Boschin\inst{6,7}
          \and
          V. Vacca\inst{1,8}
          \and
          A. Bonafede\inst{3,5}
           }

   \offprints{F. Govoni, email fgovoni@ca.astro.it}

   \institute{INAF - Osservatorio Astronomico di Cagliari,
              Loc. Poggio dei Pini, Strada 54, I--09012 Capoterra (CA), Italy
           \and
              Max-Planck-Institut f{\"u}r Astrophysik, Garching, Germany
           \and
              INAF - Istituto di Radioastronomia, 
              Via Gobetti 101, I--40129 Bologna, Italy
           \and     
              Institut f{\"u}r Astrophysik, Leopold-Franzens Universität 
              Innsbruck, Technikerstraße 25, 6020 Innsbruck, Austria 
           \and 
              Dipartimento di Astronomia, Univ. di Bologna, Via Ranzani 1, 40127 Bologna, Italy
           \and 
              Fundaci\'on Galileo Galilei - INAF, Rambla Jos\'e Ana Fern\'andez Perez 7, E-38712 Bre\~na Baja (La Palma), Canary Islands, Spain
           \and 
              Dipartimento di Astronomia, Universit\`a degli Studi di Trieste, via Tiepolo 11, I-34143 Trieste, Italy
           \and
              Dipartimento di Fisica, Universit\`a degli Studi di Cagliari, Cittadella Universitaria, I-09042 Monserrato (CA), Italy
              }

   \date{Received; accepted}

  \abstract
  {}
  {The goal of this work is to investigate the Faraday rotation measure ($RM$) of radio galaxies in hot galaxy clusters in order
   to establish a possible connection between the magnetic field strength and the gas temperature of the intracluster medium.}
  {We performed Very Large Array observations at 3.6\,cm and 6\,cm of two radio 
   galaxies located in A401 and Ophiuchus, a radio galaxy in A2142, and a radio galaxy located 
   in the background of A2065. All these galaxy clusters are characterized by high temperatures.}
   {We obtained detailed $RM$ images at an angular resolution of 3$''$ for most of the observed radio galaxies.
   The $RM$ images are patchy and reveal fine substructures of a few kpc in size. Under the assumption that the radio galaxies themselves have no effect on the measured RMs, these structures indicate that the intracluster magnetic fields fluctuate down to such small 
   scales. These new data are compared with $RM$ information present in the literature for cooler galaxy clusters.
   For a fixed projected distance from the cluster center, clusters with higher
   temperature show a higher dispersion of the $RM$ distributions ($\sigma_{RM}$), 
   mostly because of the higher gas density in these clusters.
   Although the previously known relation between the clusters X-ray surface brightness ($S_{X}$) 
   at the radio galaxy location and $\sigma_{RM}$ is confirmed, a possible connection between the $\sigma_{RM}-S_{X}$
   relation and the cluster temperature, if present, is very weak. 
   Therefore, in view of the current data, it is impossible to establish a strict link between the magnetic field strength and the gas temperature of the intracluster medium.} 
   {}


   \maketitle
%

\section{Introduction}

The knowledge of the magnetic field properties in clusters of galaxies
has significantly improved in recent years thanks to
different methods of analysis 
(see e.g. reviews by Govoni \& Feretti 2004,  Feretti \& Giovannini 2007, 
Ferrari et al. 2008, and references therein).
The presence of a magnetized plasma between an observer and a radio source 
changes the properties of the incoming polarized emission. In particular,
 the position angle of the linearly polarized radiation rotates by an 
amount that is proportional to the path integral of the magnetic fields 
along the line-of-sight times the electron density of the intervening medium, i.e. the so-called
Faraday rotation measure ($RM$).
Therefore, information on the intracluster 
magnetic fields can be obtained, in conjunction
with X-ray observations of the hot gas, through the analysis of the
$RM$ of radio galaxies in the background or in the galaxy clusters themselves.

The $RM$ studies of radio galaxies have been carried out on both 
statistical samples (e.g. Lawler \& Dennison 1982, Vall\'ee et al. 1986,
Kim et al. 1990, Kim et al. 1991, Clarke et al. 2001) 
and individual galaxy clusters and galaxy groups 
by analyzing detailed high resolution $RM$ images 
(e.g. Perley \& Taylor 1991, Taylor \& Perley 1993, Feretti et al. 1995, Feretti et al. 1999a, Feretti et al. 1999b, 
Govoni et al. 2001, Taylor et al. 2001, Eilek \& Owen 2002, 
Govoni et al. 2006, Taylor et al. 2007, Guidetti et al. 2008, Laing et al. 2008, Guidetti et al. 2010, Bonafede et al. 2010).
These data are usually consistent with central magnetic field 
strengths of a few $\mu$G, but stronger fields, with values exceeding
$\simeq$10\,$\mu$G, are derived in the inner regions of relaxed 
cooling core clusters.

For both merging and relaxed clusters, the $RM$ distribution seen
over extended radio galaxies is generally patchy, indicating that the intracluster magnetic
fields are not regularly ordered, but instead they have turbulent structures on linear
scales as low as $10~ {\mathrm kpc}$ or less.
 
In the last years, important progress has been made in the analysis
of the magnetic field structure in galaxy clusters.
In particular, it has been established that the magnetic field is likely to fluctuate 
over a wide range of spatial scales and, in a few galaxy clusters containing radio
sources with very detailed $RM$ images, the magnetic field power spectrum has 
been estimated
(e.g. En{\ss}lin \& Vogt 2003, Vogt \& En{\ss}lin 2003, Murgia et al. 2004,  
Vogt \& En{\ss}lin 2005, Govoni et al. 2006, Guidetti et al. 2008, 
Laing et al. 2008, Guidetti et al. 2010, Bonafede et al. 2010).

An increasing attention is given in the literature 
to the connection between the magnetic field strength and the cluster 
gas density and temperature (e.g. Kunz et al. 2010).
SPH simulations (Dolag et al. 1999, 2002, 2005a), indicated 
that the magnetic field intensity 
declines from the center of the cluster outward with a 
rough dependence on the thermal gas density. 
Such magnetic field profiles were confirmed by other simulations 
based on different numerical techniques 
(e.g. Br{\"u}ggen et al. 2005, Dubois \& Teyssier 2008, Collins et al. 2009). 
SPH simulations also predicted a very steep correlation of the 
mean magnetic field with the cluster temperature, 
e.g. $B\propto T^2$ (Dolag et al. 1999). However, there is also the possibility that the magnetic field 
strength is only mildly dependent on the cluster temperature 
as recently pointed out by Donnert et al. (2009). 
Kunz et al. (2010) predicted a dependence 
of the magnetic field strength on the cluster temperature in cool-core clusters 
($B\propto n_e^{1/2}T^{3/4}$ in cool-core clusters, $B\propto n_e^{1/2}$ in isothermal clusters).

Dolag et al. (2001) showed that the correlation between two
observable parameters, the dispersion of the $RM$ distribution ($\sigma_{RM}$)
and the cluster X-ray surface brightness in the source location ($S_{X}$),
is expected to reflect a correlation between the cluster magnetic
field and the gas density. Indeed, measuring the slope 
of the $\sigma_{RM}-S_{X}$ correlation it is possible to
infer the trend of the magnetic field versus the gas density.  
In addition, as predicted by cosmological magneto-hydrodynamic 
simulations (Dolag et al. 1999), galaxy clusters 
should have different central magnetic field strengths
depending on their dynamical state and hence temperature, leading to
an offset of this correlation for hot clusters.

In this work we present sensitive polarimetric Very Large Array (VLA) 
radio observations at multiple frequencies of polarized radio sources embedded
in a set of hot, nearby clusters of galaxies.
These data are compared with $RM$ data taken from literature, in order to
investigate a possible connection between the magnetic field strength and the gas 
temperature of the intracluster medium.
Moreover, a detailed analysis of the Faraday rotation of the most interesting sources
presented in this work will allow us, in a forthcoming paper, 
to determine the cluster magnetic field properties, by following the approach proposed 
in Murgia et al. (2004) and applied in the cases of
A2255 (Govoni et al. 2006), A2382 (Guidetti et al. 2008), and Coma (Bonafede et al. 2010).

The paper is organized as follows: 
in Sect. 2 we describe the selection criteria of the targets,
we illustrate the radio observations, and the data reduction. 
In Sect. 3 we show the results of the total intensity and polarization properties 
of the radio galaxies. In Sect. 4 we show the $RM$ images.
In Sect. 5 we investigate the magnetic field - gas temperature connection by comparing these new $RM$ data 
and some other high quality data taken from 
literature with the cluster temperature. Finally, in Sect. 6 
we draw the conclusions.

Throughout this paper we assume a $\Lambda$CDM cosmology with
$H_0= 71~\mathrm{ km\, s^{-1}\, Mpc^{-1}}$,
$\Omega_m=0.27$, and $\Omega_{\Lambda}=0.73$.


\section{VLA observations and data reduction}

\begin{table*}
\caption{List of hot, nearby clusters of galaxies hosting the selected targets.
}
\begin{center}
\begin{tabular} {ccccc} 
\hline
Cluster   & z & kpc/$''$ &  T     & Ref. \\
          &          &   & [keV]  &      \\
\hline
A401      & 0.074 & 1.39 & $8.3\pm0.5$      & Markevitch (1998) \\
A2142     & 0.091 & 1.67 & $8.8\pm0.6$      & Markevitch (1998)     \\
A2065     & 0.073 & 1.37 & $5.4\pm0.3$      & Markevitch (1998)    \\
Ophiuchus & 0.028  & 0.55 & $10.26\pm0.32$  & Fukazawa et al. (1998) \\
\hline
\multicolumn{5}{l}{\scriptsize Col. 1: Cluster Name; Col. 2: Redshift; Col. 3: Angular to linear conversion;}\\ 
\multicolumn{5}{l}{\scriptsize Col. 4: Cluster temperature measured with ASCA; Col. 5: Temperature reference.}\\ 
\end{tabular}
\label{tab1}
\end{center}
\end{table*}

\begin{table*}
\caption{Details of the VLA observations.}
\begin{center}
\begin{tabular} {lccccccc} 
\hline
Source   &  R.A.         &  Decl.         &  Frequency & Bandwidth &  Conf. &  Time   & Date \\
         &  J2000      &  J2000       &  ${\mathrm MHz}$       & ${\mathrm MHz}$       &        &  hours  &      \\
\hline
A401A    & 02 58 32   & $+$13 34 16   &  4885/4535   &  50  &  B  & 1.2, 1.5 & 2002-Jul-13, 2003-Dec-22  \\
         &            &               &  4885/4535   &  50  &  C  & 1.2      & 2002-Nov-04 \\
         &            &               &  8085/8465   &  50  &  B  & 1.2      & 2002-Jul-13\\
         &            &               &  8085/8465   &  50  &  C  & 1.2, 1.4 & 2002-Nov-04, 2004-Apr-16 \\
A401B    & 02 59 15   & $+$13 27 36   &  4885/4535   &  50  &  B  & 1.2, 1.5 & 2002-Jul-13, 2003-Dec-22 \\
         &            &               &  4885/4535   &  50  &  C  & 1.2      & 2002-Nov-04\\
         &            &               &  8085/8465   &  50  &  B  & 1.2      & 2002-Jul-13\\
         &            &               &  8085/8465   &  50  &  C  & 1.2, 1.4 & 2002-Nov-04, 2004-Apr-16\\
A2142A   & 15 58 13   & $+$27 16 24   &  4885/4535   &  50  &  B  & 3.3      & 2003-Dec-23\\
         &            &               &  8085/8465   &  50  &  C  & 3.1      & 2004-Apr-13\\
A2065A   & 15 22 46   & $+$27 55 26   &  4885/4535   &  50  &  B  & 3.3      & 2003-Dec-23\\
         &            &               &  8085/8465   &  50  &  C  & 3.1      & 2004-Apr-13\\
OPHIB    & 17 11 54   & $-$23 09 23   &  4885/4835   &  50  & BnA & 1.6      & 2003-Oct-07\\
         &            &               &  8085/8465   &  50  & CnB & 1.6      & 2004-Feb-08\\
OPHIC    & 17 12 09   & $-$23 28 43   &  4885/4835   &  50  & BnA & 1.6      & 2003-Oct-07\\
         &            &               &  8085/8465   &  50  & CnB & 1.6      & 2004-Feb-08\\
\hline
\multicolumn{8}{l}{\scriptsize Col. 1: Source; Col. 2,3: Pointing position (R.A., Decl.); Col. 4: Observing frequency;}\\
\multicolumn{8}{l}{\scriptsize Col. 5: Observing Bandwidth; Col. 6: VLA configuration;}\\
\multicolumn{8}{l}{\scriptsize Col. 7: Time on source; Col. 8: Dates of observation.}\\
\end{tabular}
\label{tab2}
\end{center}
\end{table*}

We selected a small sample of radio galaxies embedded in a set of nearby, 
high temperature galaxy clusters. 
Starting from the X-ray flux limited sample by Edge et al. (1990),
we considered all the clusters with a temperature $T\geq 8~ {\mathrm keV}$, a redshift 
$z<0.1$, and declination $\delta > -40^\circ$ (to be observed by the VLA). 
Among the resulting nine clusters (Ophiuchus, Coma, A2319, A754, A2142, A401, A644, A2065, A2255) we skipped
Coma and A2255 because they were already analyzed in other 
$RM$ projects (Feretti et al. 1995, Govoni et al. 2006, Bonafede et al. 2010), and 
A754 because of its complex X-ray morphology. 
Of the remaining clusters, we searched in the NRAO VLA Sky Survey (NVSS; Condon et al. 1998) 
for bright and extended radio galaxies. 
We selected all the radio galaxies located within 15$'$ from the clusters center
showing in the NVSS an angular extension ${\mathrm LAS}>1'$ 
and a flux density at $1.4~ {\mathrm GHz} \geq60~ {\mathrm mJy}$. There were no radio galaxies
in A644 that meet the above selection criteria. In A2319 a radio galaxy  was selected, 
but the data were corrupted and not useful for our analysis.
Thus, the final list of the new observed clusters, together with their redshift, 
the angular to linear conversion, the ASCA cluster 
temperature, and the corresponding temperature reference is given in Table \ref{tab1}.
We note that although A2065 is reported by Edge et al. (1990) and David et al. (1993) with a 
high temperature, other X-ray investigation found for 
this cluster a temperature $T<8~ {\mathrm keV}$ (e.g. Markevitch 1998).

Here we present the results of multifrequency, polarimetric observations of these sources obtained with the VLA.
The details of the observations are provided in Table \ref{tab2}.
The sources were observed at $4885~ {\mathrm MHz}$ and $4535~ {\mathrm MHz}$ 
($4835~ {\mathrm MHz}$ for Ophiuchus)
within the $6\, {\mathrm cm}$ band and at $8085~ {\mathrm MHz}$ and $8465~ {\mathrm MHz}$ within the $3.6\, {\mathrm cm}$ band.
A401 was observed within each band in both the B and C 
configurations. A2142 and A2065 were observed in the B configuration within the $6\, {\mathrm cm}$ band and 
in the C configuration within the $3.6\, {\mathrm cm}$ band.  
Given its southern declination, 
 Ophiuchus was observed with  the VLA in the hybrid configurations BnA and CnB at $6\, {\mathrm cm}$  and $3.6\, {\mathrm cm}$, respectively.
All observations were made with a bandwidth of $50~ {\mathrm MHz}$.

Calibration and imaging were performed with the NRAO Astronomical 
Image Processing System (AIPS) following the standard procedure:
Fourier-Transform, Clean and
Restore. Self-calibration was applied to remove residual 
phase variations. In the case of A401 the ($u$,$v$) 
data at the same frequencies
but from different configurations were first handled separately and then
combined.

Images of the Stokes parameters $I$, $U$, and $Q$ have been obtained for 
each frequency separately. 
Images of polarized intensity $P=(Q^2+U^2)^{1/2}$ (corrected for the positive bias), 
fractional polarization $FPOL=P/I$ and position angle of
polarization $\Psi=0.5\tan^{-1}(U/Q)$ were derived from the $I$, $Q$, 
and $U$ images.

\section{Total intensity and polarization properties}
In the following we give a brief description of total intensity and polarization properties
of the individual sources.

For each galaxy cluster the radio contours obtained from the NVSS 
overlaid on the ROSAT HRI X-ray images are shown (Figs. 1, 4, 6, 8).
The NVSS images are at $1.4~ {\mathrm GHz}$ and have an angular 
resolution of $45''$.
The X-ray images are in the $0.1-2.4~ {\mathrm keV}$ band and have been
convolved with a Gaussian with $\sigma=16''$.
High resolution images of the radio galaxies, in the $6\, {\mathrm cm}$ band,
overlaid on optical red plate are inset.
Every radio galaxy shows an optical
counterpart in the DSS2\footnote{htpp://archive.eso.org/dss/dss}, except for A2065A
which is likely a cluster background source.

In addition, for each source total intensity and polarization images at $3.6\, {\mathrm cm}$ and $6\, {\mathrm cm}$, convolved with the same beam of 
$3''$, are shown (Figs. 2, 3, 5, 7, 9, 10). 
Contours represent total intensity while vectors represent the orientation of the projected E-field and are proportional
in length to the fractional polarization. 
In the fractional polarization images we considered as valid pixels those where
the fractional polarization FPOL was above $3\sigma_{\mathrm FPOL}$. 
Sample plots of the trend of the fractional polarization as a function of the observing frequencies, 
calculated in different regions of the sources are also shown. 

We note that in some sources we found some unpolarized areas.
This is likely due to a combination of sensitivity limits, 
small-scale tangled fields, and depolarization.  
The calculation of the average fractional polarization 
may be biased if these depolarized regions are ignored.  
To reduce any possible bias, we calculated the 
fractional polarization over the sources in all the pixels where the
total intensity signal is above 5$\sigma_{\mathrm I}$ at all frequencies.

The relevant parameters of the $I$, $Q$, $U$ images, convolved with a beam
of $3''$, are listed in Table \ref{mappeI}. The total flux densities,
are also reported. The total flux densities were calculated, after the primary beam correction, 
by integrating the total intensity surface brightness, down to the noise level.

\begin{table*}
\caption{Information on total intensity and polarization images at 
$6\, {\mathrm cm}$ and $3.6\, {\mathrm cm}$}
\begin{center}
\begin{tabular} {lcccccr} 
\hline
Source   &  $\lambda$   &  Beam         &  $\sigma$(I)$^{*}$ & $\sigma$(Q)   &  $\sigma$(U) & Flux density \\
         &  (cm)        &  (\arcsec)    &  (mJy/beam)  & (mJy/beam)    &  (mJy/beam)  &     (mJy)     \\
\hline
A401A    &  3.6         &3.0$\times$3.0   & 0.021 &  0.018  & 0.018   &  95.0$\pm$3.0   \\
         &  6           & $''$            & 0.031 &  0.027  & 0.028   & 159.0$\pm$5.0  \\
A401B    &  3.6         & $''$            & 0.025 &  0.025  & 0.024   &  36.0$\pm$1.0  \\
         &  6           & $''$            & 0.033 &  0.028  & 0.029   &  55.0$\pm$2.0  \\
A2142A   &  3.6         & $''$            & 0.017 &  0.015  & 0.016   &  22.0$\pm$0.7  \\
         &  6           & $''$            & 0.029 &  0.030  & 0.028   &  37.0$\pm$1.0  \\
A2065A   &  3.6         & $''$            & 0.018 &  0.015  & 0.015   &  19.0$\pm$0.6   \\
         &  6           & $''$            & 0.038 &  0.030  & 0.029   &  32.0$\pm$1.0   \\
OPHIB    &  3.6         & $''$            & 0.044 &  0.038  & 0.036   &  46.0$\pm$2.0  \\
         &  6           & $''$            & 0.115 &  0.095  & 0.086   &  58.0$\pm$2.0  \\
OPHIC    &  3.6         & $''$            & 0.036 &  0.036  & 0.039   &  13.0$\pm$0.4  \\
         &  6           & $''$            & 0.089 &  0.092  & 0.093   &  23.0$\pm$0.7   \\
\hline
\multicolumn{7}{l}{\scriptsize Col. 1: Source; Col. 2: Observation wavelength; Col. 3: Beam;}\\ 
\multicolumn{7}{l}{\scriptsize Col. 4, 5, 6: RMS noise of the $I$, $Q$, $U$ images; Col. 7: Flux density.}\\
\multicolumn{7}{l}{\scriptsize $^{*}$ Note that in this table we indicate the values of $\sigma$(I), $\sigma$(Q),  
$\sigma$(U), and Flux density for the frequencies}\\
 \multicolumn{7}{l}{\scriptsize 8465 MHz and 4535 MHz (4835 MHz for OPHIB and OPHIC) at 3.6 and $6\, {\mathrm cm}$ respectively.}\\
\end{tabular}
\label{mappeI}
\end{center}
\end{table*}

\subsection{Abell 401}

\begin{figure*}
\vspace{9 cm}
\includegraphics{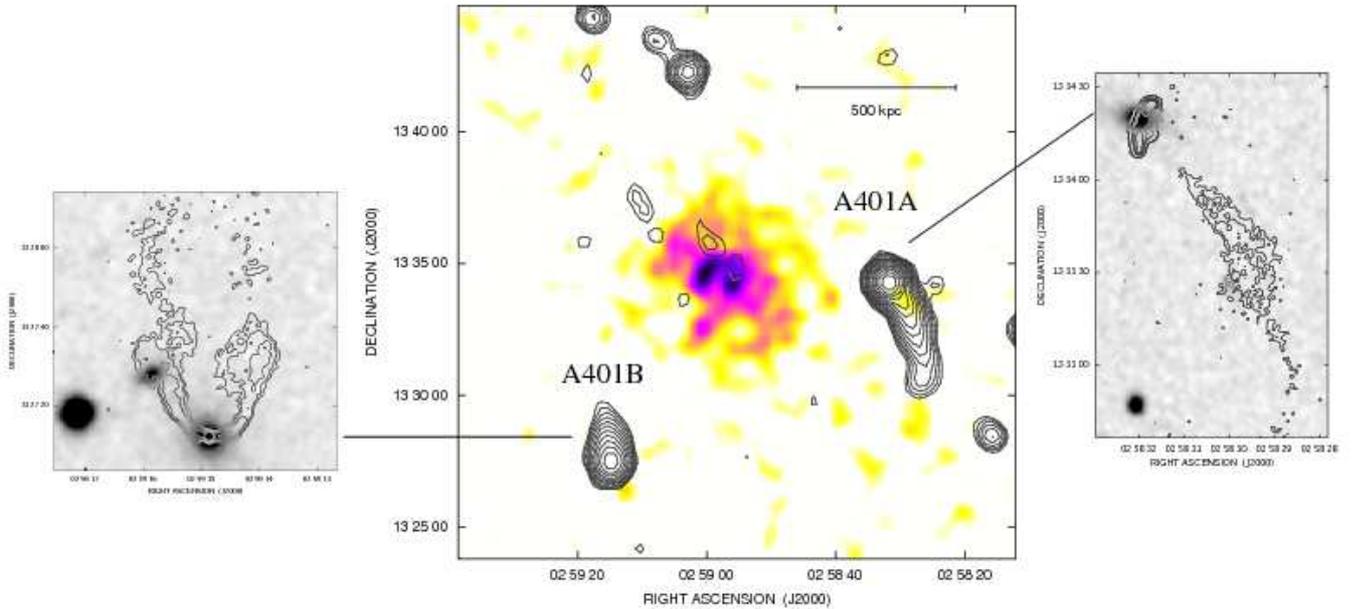}
\caption[]{
1.4 GHz radio image of A401 obtained from the NVSS (contours) 
overlaid on the ROSAT HRI X-ray image (colors) in the 
$0.1-2.4~ {\mathrm keV}$ band.
The NVSS image has an angular resolution of 45\arcsec.
The first radio contour is drawn at 1.5 mJy/beam and the rest are spaced by a
factor of $\sqrt 2$. 
The high resolution observations at $6\, {\mathrm cm}$ (contours) 
overlaid on the DSS2 red plate (grey-scale) of the two radio galaxies are inset. 
A401A : Contour levels = 0.05 0.1 1 2 4 6 mJy/beam.
Angular resolution = $1.67''\times 1.42''$ (PA=$-$45.2).
Sensitivity = 0.015 mJy/beam. 
A401B : Contour levels = 0.05 0.1 0.2 0.5 1 mJy/beam.
Angular resolution = $1.52''\times 1.38''$ (PA=$-$19.8).
Sensitivity = 0.016 mJy/beam. 
}
\label{A401}
\end{figure*}

\begin{figure*}
\vspace{13cm}
\includegraphics{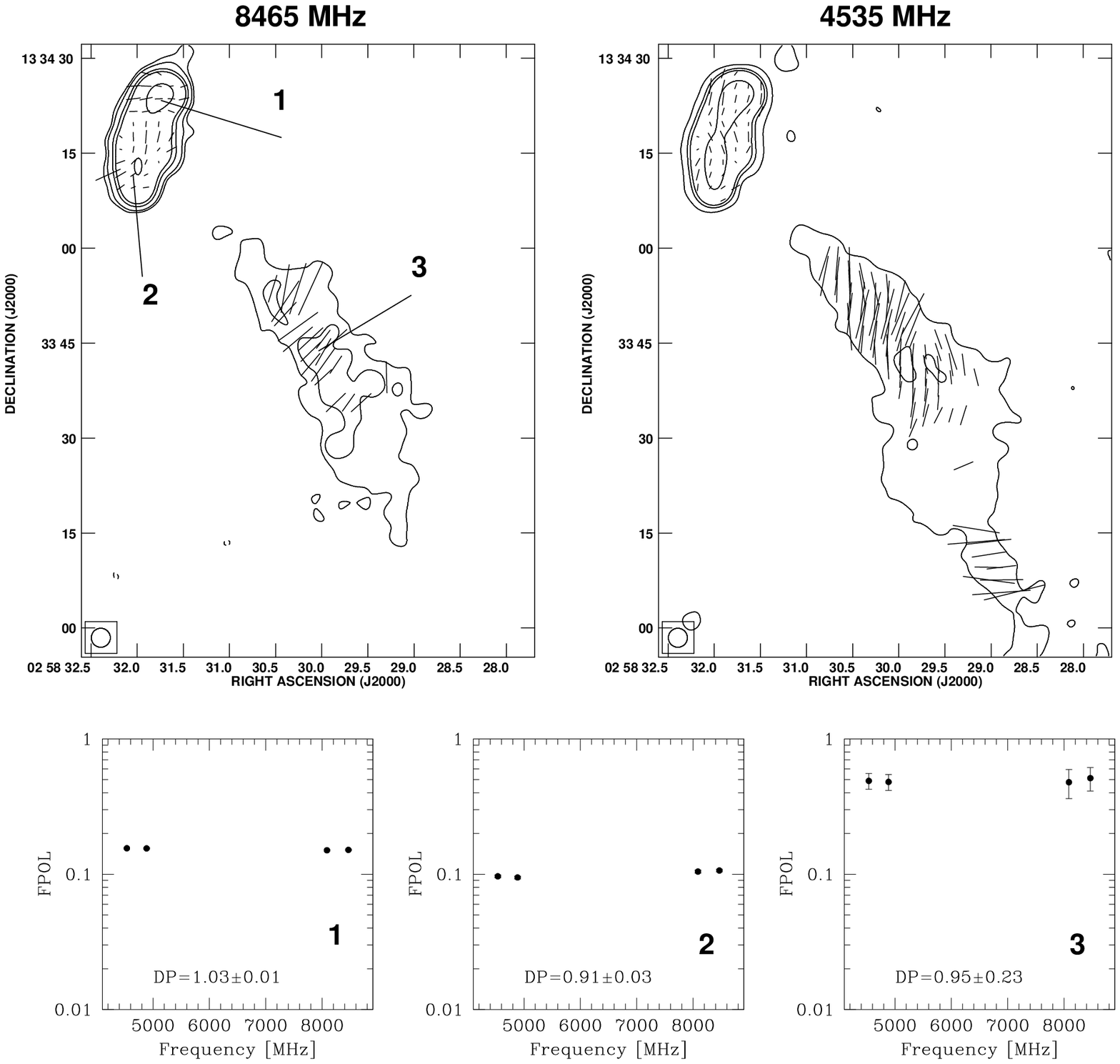}
\caption[]{
Source A401A.
Top, left: Total intensity contours and polarization vectors 
at $3.6\, {\mathrm cm}$ (8465 ${\mathrm MHz}$). Contour levels
are drawn at: -0.07 0.07 0.15 0.5 1 and 10 mJy/beam.
Top, right: Total intensity contours and 
polarization vectors at $6\, {\mathrm cm}$ (4535 ${\mathrm MHz}$). 
Contour levels are drawn at: -0.1 0.1 0.5 1 and 10 mJy/beam.
The angular resolution is $3.0''\times 3.0''$. The lines give the
orientation of the electric vector position angle (E-field)
and are proportional in length to the fractional polarization 
($1''\simeq10$\%).
Bottom: Trend of the fractional polarization as a function of the observing frequencies, 
at different source locations. The depolarization $DP$ has been calculated as the ratio of the fractional polarization between the two extreme frequencies.
}
\label{a401a}
\end{figure*}

\begin{figure*}
\vspace{12cm}
\includegraphics{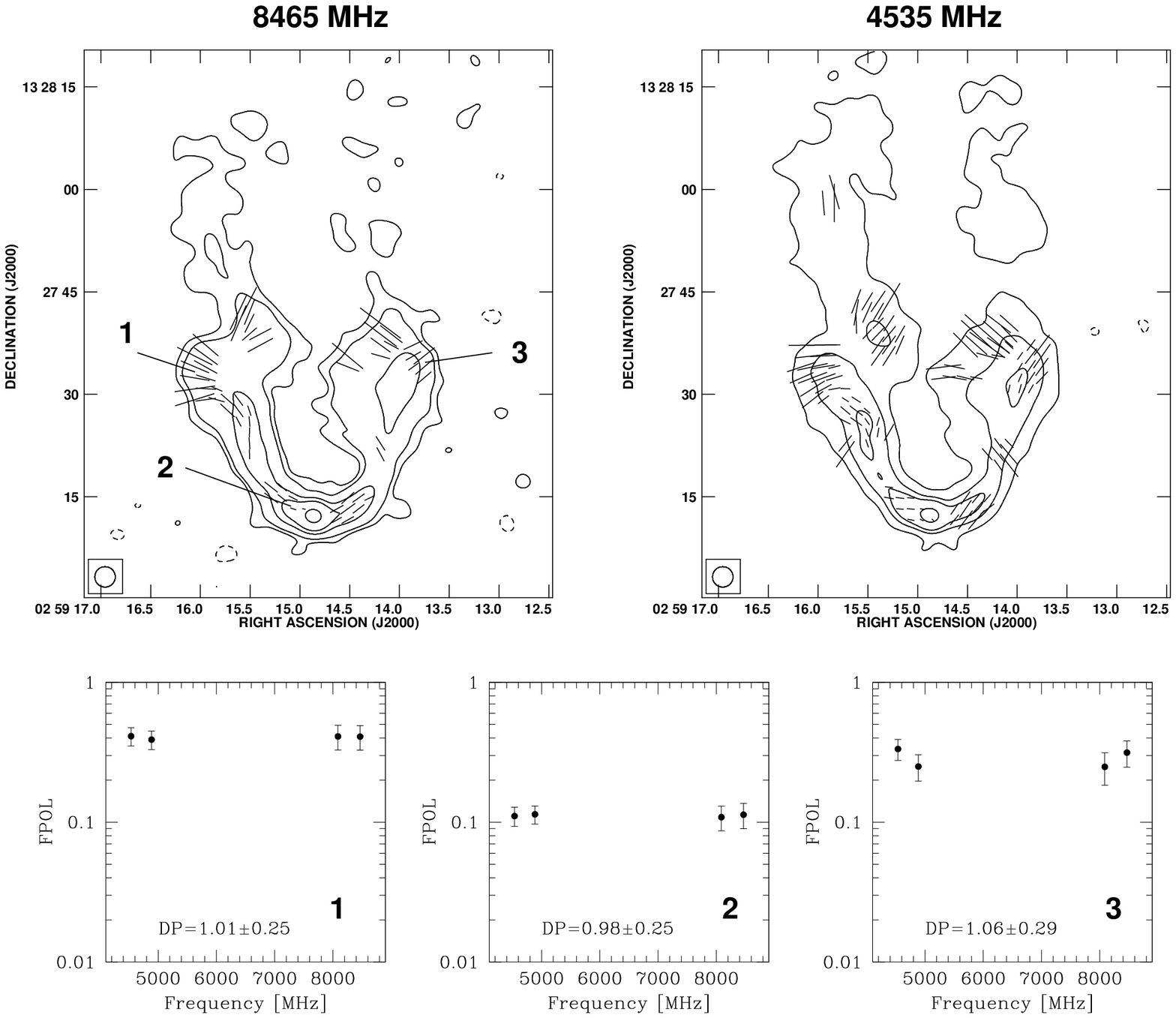}
\caption[]{
Source A401B.
Top, left: Total intensity contours and polarization vectors 
at $3.6\, {\mathrm cm}$ (8465 ${\mathrm MHz}$). Contour levels
are drawn at: -0.07 0.07 0.15 0.5 1 and 3 mJy/beam.
Top, right: Total intensity contours and 
polarization vectors at $6\, {\mathrm cm}$ (4535 ${\mathrm MHz}$).
Contour levels are drawn at: -0.1 0.1 0.5 1 and 3 mJy/beam. 
The angular resolution is $3.0''\times 3.0''$. 
The lines give the orientation of the electric vector position angle (E-field)
and are proportional in length to the fractional polarization 
($1''\simeq10$\%)
Bottom: Trend of the fractional polarization as a function of the observing frequencies, 
at different source locations.
The depolarization $DP$ has been calculated as the ratio of the fractional polarization between the two extreme frequencies.
}
\label{a401b}
\end{figure*}

The ROSAT HRI X-ray image presented in Fig. \ref{A401} 
shows the X-ray emission in the $0.1-2.4~ {\mathrm keV}$ band of the cluster of 
galaxies A401. 
In the southwest, at a projected distance of about $3~ {\mathrm Mpc}$,
not shown in the field of view of the image, lies the A399 system.
The X-ray excess and the slight temperature increase in the region 
between the two clusters indicate a physical link between 
this merging pair of clusters (e.g. Fujita et al. 1996,
Fabian et al. 1997, Markevitch et al. 1998, Sakelliou \& Ponmann 2004, 
Bourdin \& Mazzotta 2008). 
Recent XMM X-ray analyses (Sakelliou \& Ponman 2004, Bourdin \& Mazzotta 2008)
pointed out that neither of the two clusters contain a cooling core and 
their central regions are nearly isothermal, with some small-scale inhomogeneities.
However, the reasonably relaxed morphology of the clusters and the absence
of major temperature anomalies argue against models in which A399 and A401 have already
experienced a close encounter (see also Fujita et al. 2008a).

A401 is known to contain a faint radio halo around the cD galaxy at the cluster center
(Harris et al. 1980, Roland et al. 1981, Giovannini et al. 1999, Bacchi et al. 2003).
We recently found (Murgia et al. 2010a) that the central region of A399 is also permeated by a diffuse low-surface brightness 
radio emission which we classified as radio halo. Indeed, given their proximity to each other, 
the interacting pair of clusters of galaxies A401 and A399 can be considered as the first example 
of double radio halo system.

The periphery of A401 hosts a number of tailed radio galaxies.
The head-tail morphology is probably caused by a sweeping 
of the radio emitting material
due to the relative motion between the gas and the radio galaxy.
In this work we have analyzed those labeled with A401A and A401B 
in Fig. \ref{A401}.

The distribution of spectroscopic redshifts 
near the pair of clusters A399 and A401
shows only one peak. This is because the velocity dispersion for
individual clusters is larger than the apparent velocity separation 
between A401 and A399 (Hill \& Oegerle 1993, Oegerle \& Hill 1994, 
Yuan et al. 2005).
On the basis of their radial velocity and their projected distance from A401, 
the two galaxies associated with the radio sources A401A and A401B
are classified as cluster members.
 
A401A is located to the west of the
cluster center at a projected distance
of about 5.8$'$. 
The high resolution image at $6\, {\mathrm cm}$ reveals a peculiar morphology. 
A compact slightly distorted double structure of about  
20$''$ in size whose optical counterpart is located at 
R.A.(J2000)=02h58m32.0s, Decl.(J2000)=13\degrees 34\arcmin 20\arcsec,
and a low surface brightness tail of more than 2$'$ in size
with a morphology which appears not directly connected
to the compact emission.
The total flux density of the radio galaxy, both at 
$6\, {\mathrm cm}$ and  $3.6\, {\mathrm cm}$, is indicated in Table 3.
The compact component has a flux density of 
$134\pm 4~ {\mathrm mJy}$ and $86\pm 3~ {\mathrm mJy}$ at 
$6\, {\mathrm cm}$ and  $3.6\, {\mathrm cm}$ respectively leading
to a mean spectral index\footnote{S($\nu$)$\propto \nu^{- \alpha}$ with $\alpha$=spectral index} 
 $\alpha$ $\simeq$ 0.7$\pm$0.1, typical for radio galaxies.
In the tail the spectral index is very steep ($\alpha$ $\simeq$ 1.6$\pm$0.1). In fact the 
flux density is $25 \pm 1~ {\mathrm mJy}$ and 
$9.0 \pm 0.3~ {\mathrm mJy}$ at $6\, {\mathrm cm}$ and  $3.6\, {\mathrm cm}$
respectively.

The head tail A401B is located
at a projected distance of about 8.5\arcmin~to the southeast of the
cluster center with the head pointed versus the 
south of the cluster and the tail elongated versus the north.
The total flux density of the radio galaxy, both at 
$6\, {\mathrm cm}$ and  $3.6\, {\mathrm cm}$, is indicated in Table 3.
The high resolution image at $6\, {\mathrm cm}$ shows the core, in the position
R.A.(J2000)=02h59m14.9s, Decl.(J2000)=13\degrees 27\arcmin 12\arcsec,
coincident with a galaxy. At $6\, {\mathrm cm}$ the source is about 1\arcmin~in size. 

Figs. \ref{a401a} and \ref{a401b} show the polarization 
images at 8465 ${\mathrm MHz}$ and 4535 ${\mathrm MHz}$, with an angular resolution of $3''\times3''$.

The polarization is clearly detected at both frequencies in the bright compact double structure of A401A. The tail of A401A has a steep spectrum, consequently, the total intensity is drastically reduced at 8465 ${\mathrm MHz}$ and the polarized signal is detected only in 
the brightest regions of the tail.
By considering all the pixels where the total intensity 
signal is above 5$\sigma_{\mathrm I}$ at
all frequencies, the compact double structure of A401A has a mean fractional polarization of
$12\pm3$\% both at 8465 and 4535 ${\mathrm MHz}$.  In the tail the fractional polarization
is higher than in the compact component reaching a mean fractional polarization of $36\pm5$\% both 
at 8465 and 4535 ${\mathrm MHz}$.
In the bottom panels of Fig. \ref{a401a} we show the trend of the fractional 
polarization as a function of the observing frequencies, at different source locations. 
The fractional polarization does not decrease significantly in the analyzed 
range of wavelengths.

A401B is polarized at both frequencies up to a distance of about 30$''$.
The mean fractional polarization is $20\pm3$\% at 8465 and  $18\pm3$\% at 4535 ${\mathrm MHz}$.
Therefore, the depolarization is negligible.
This is also confirmed in the bottom panels of Fig. \ref{a401b} where we show the trend of the 
fraction polarization as a function of the observing frequencies, calculated in three different locations 
of the source.

\subsection{Abell 2142}

\begin{figure*}
\vspace{9cm}
\includegraphics{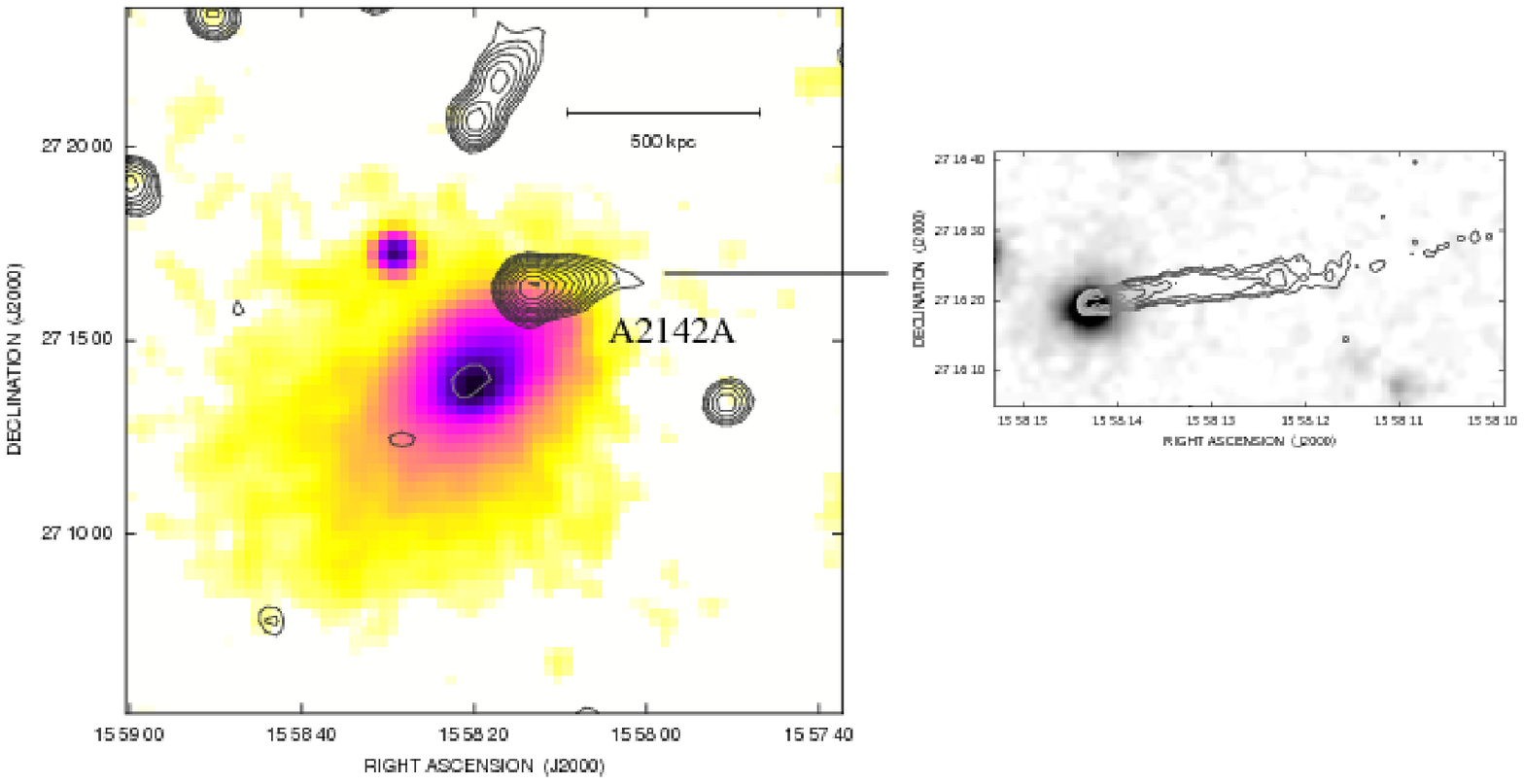}
\caption[]{
1.4 GHz radio image of A2142 obtained from the NVSS (contours) 
overlaid on the ROSAT HRI X-ray image (colors) in the $0.1-2.4~ {\mathrm keV}$ band.
The NVSS image has an angular resolution of 45\arcsec.
The first radio contour is drawn at 1.5 mJy/beam and the rest are spaced by a
factor of $\sqrt 2$.
The high resolution observation at $6\, {\mathrm cm}$ (contours) overlaid on the DSS2 
red plate (grey-scale) of the radio galaxy A2142A is inset. 
Contour levels = 0.05 0.1 0.2 0.5 1 2 mJy/beam.
Angular resolution = $1.39'' \times 1.18''$ (PA=$-$57.5).
Sensitivity = 0.014 mJy/beam. 
}
\label{A2142}
\end{figure*}

\begin{figure*}
\vspace{10cm}
\includegraphics{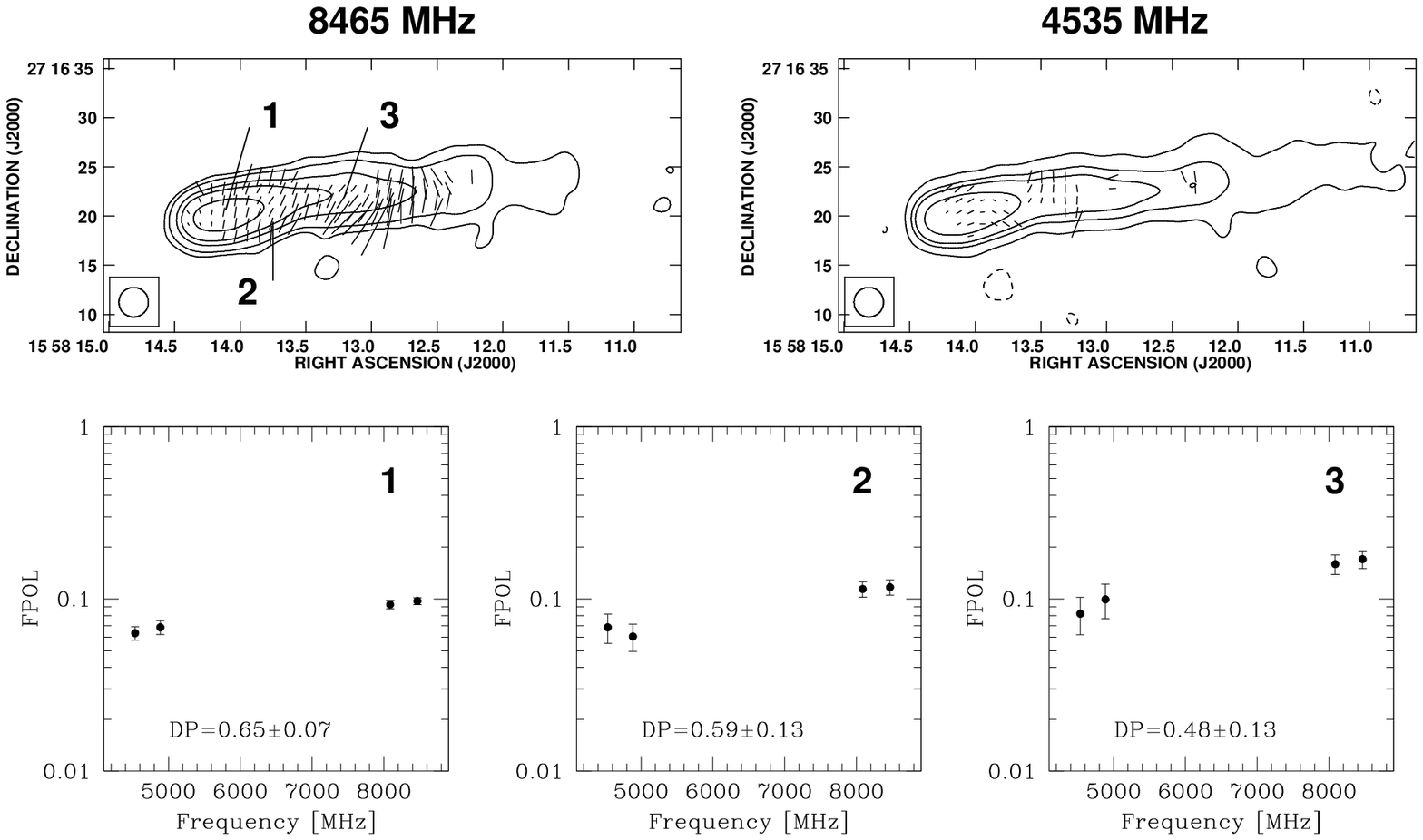}
\caption[]{
Source A2142A.
Top, left: Total intensity contours and polarization vectors 
at $3.6\, {\mathrm cm}$ (8465 ${\mathrm MHz}$). Contour levels
are drawn at: -0.06 0.06 0.15 0.5 and 3 mJy/beam.
Top, right: Total intensity contours and 
polarization vectors at $6\, {\mathrm cm}$ (4535 ${\mathrm MHz}$). 
Contour levels are drawn at: -0.1 0.1 0.5 1 and 3 mJy/beam.
The angular resolution is $3.0''\times 3.0''$. The lines give the
orientation of the electric vector position angle (E-field)
and are proportional in length to the fractional polarization 
($1''\simeq10$\%).
Bottom: Trend of the fractional polarization as a function of the observing frequencies, 
at different source locations.
The depolarization $DP$ has been calculated as the ratio of the fractional polarization between the two extreme frequencies.
}
\label{a2142a}
\end{figure*}

The ROSAT HRI X-ray image presented in Fig. \ref{A2142} 
shows the X-ray emission of the cooling flow cluster A2142.
The cooling flow manifests itself with a cool X-ray peak 
as well as strong evidence for a centrally
enhanced metal abundance (e.g. White et al. 1994, Peres et al. 1998).
On the other hand optical (Oegerle et al. 1995) and X-ray 
(Henry \& Briel 1996, Buote \& Tsai 1996, Pierre \& Starck 1998)
features, such as the presence of an off-center, small scale structures 
and azimuthally asymmetric temperature variations, are indications of a merger. 
Chandra data show that the gas temperature in the central region of A2142
(Markevitch et al. 2000) is non-isothermal.
It appears that the central cooling flow has been disturbed but not 
destroyed by the merger. 

The presence of a halo in this cluster was suggested by
Harris et al. (1977). Giovannini \& Feretti (2000) confirmed 
the presence of diffuse emission,
located around the brightest cluster galaxy. This source, however,
is much smaller than radio halos commonly
found in merging clusters and it may be a mini-halo,
like those found at the center of relaxed systems (see e.g. Gitti et al. 2007, 
Govoni et al. 2009). 

In this work we have analyzed the radio galaxy labeled with A2142A
in Fig. \ref{A2142}.
This narrow head tail radio galaxy has the head
at a projected distance of about 2.7\arcmin~to the northwest of the
cluster X-ray center and the tail elongated versus the west.
The high resolution image at $6\, {\mathrm cm}$ shows the head, at the position
R.A.(J2000)=15h58m14.3s, Decl.(J2000)=27\degrees16\arcmin19.5\arcsec,
coincident with a galaxy. At $6\, {\mathrm cm}$ the source is more than 40\arcsec~in size. 
In the source we do not distinguish the two jets. A possibility is that the 
source is narrow because of projection effects, i.e. the radio jets are 
seen edge-on.

Fig. \ref{a2142a} shows the polarization 
images at 8465 ${\mathrm MHz}$ and 4535 ${\mathrm MHz}$, with an angular resolution of $3''\times3''$.
At 8465 ${\mathrm MHz}$ the polarization is significant (above 3$\sigma_{\mathrm FPOL}$) 
up to a distance of about 30$''$, while at 4535 ${\mathrm MHz}$ the polarization signal
appear below the noise level in some places along the tail. 
By considering all the pixels where the total intensity signal is 
above 5$\sigma_{\mathrm I}$ at all frequencies, the mean fractional 
polarization is $16\pm4$\% at 8465 and  $10\pm3$\% at 4535 ${\mathrm MHz}$.
The source is therefore depolarized    
$<DP>=FPOL_{4535 {\mathrm MHz}}/FPOL_{8465 {\mathrm MHz}}\simeq 0.6$.
The depolarization of the signal is also evident in the bottom panels 
of Fig. \ref{a2142a}, where we show the trend of the fractional 
polarization as a function of the observing frequencies, at different source locations.  

\subsection{Abell 2065}

\begin{figure*}
\vspace{9cm}
\includegraphics{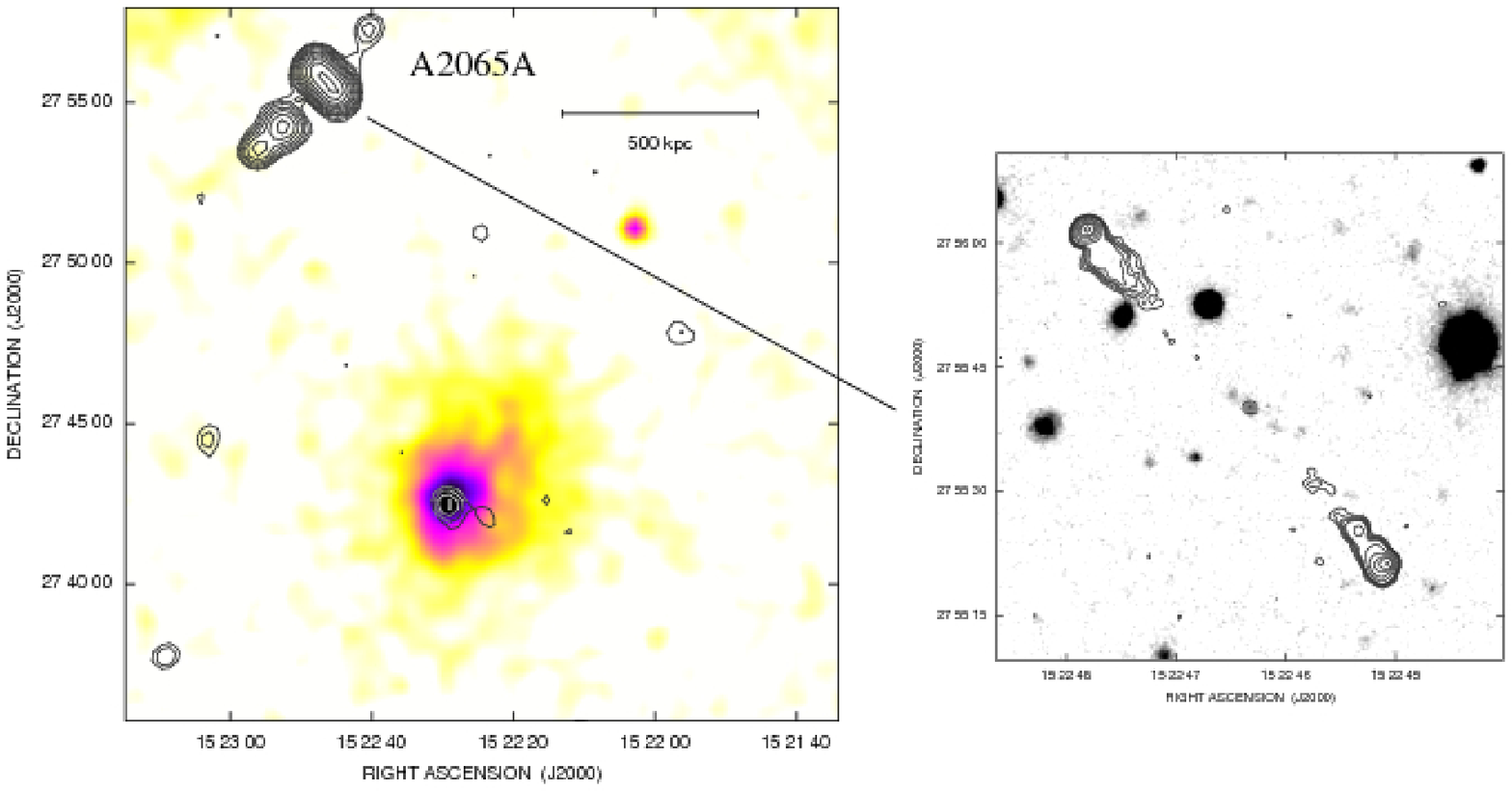}
\caption[]{
1.4 GHz radio image of A2065 obtained from the NVSS (contours) 
overlaid on the ROSAT HRI X-ray image (colors) in the $0.1-2.4~ {\mathrm keV}$
 band.
The NVSS image has an angular resolution of 45\arcsec.
The first radio contour is drawn at 1.5 mJy/beam and the rest are spaced by a
factor of $\sqrt 2$. 
The high resolution observation at $6\, {\mathrm cm}$ (contours) overlaid on a TNG R-band image (grey-scale)
of the radio galaxy A2065A is inset. Contour levels = 0.05 0.08 0.12 0.15 0.2 0.5 1 2 3 5 mJy/beam.
Angular resolution = $1.24'' \times 1.18''$  (PA=$-$45.5).
Sensitivity = 0.014 mJy/beam. 
}
\label{A2065}
\end{figure*}

\begin{figure*}
\vspace{13cm}
\includegraphics{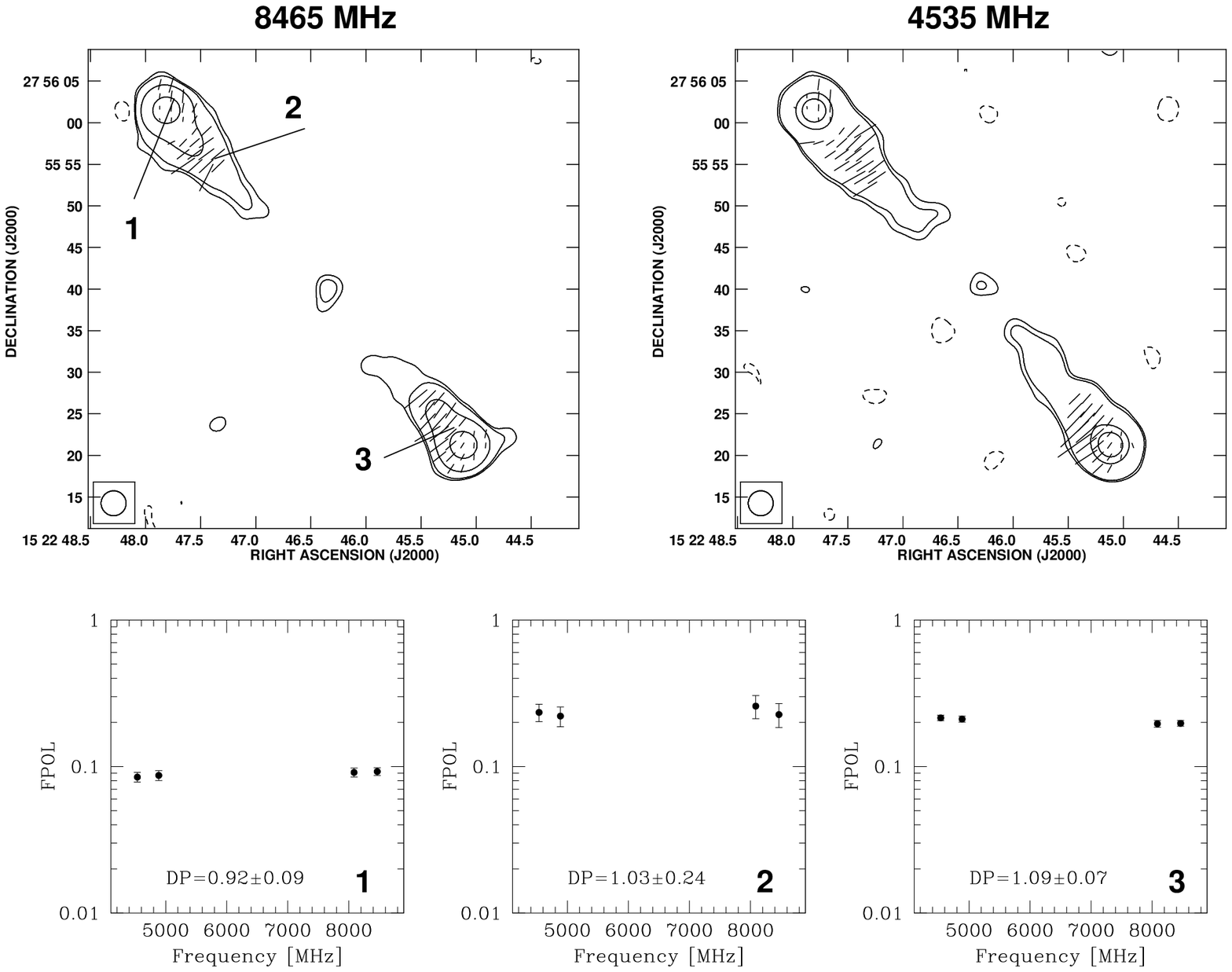}
\caption[]{
Source A2065A. 
Top, left: Total intensity contours and polarization vectors 
at $3.6\, {\mathrm cm}$ (8465 ${\mathrm MHz}$). Contour levels
are drawn at: -0.06 0.06 0.1 0.5 and 3 mJy/beam.
Top, right: Total intensity contours and 
polarization vectors at $6\, {\mathrm cm}$ (4535 ${\mathrm MHz}$). 
Contour levels are drawn at: -0.1 0.1 0.15 3 and 6 mJy/beam.
The angular resolution is $3.0''\times 3.0''$. The lines give the
orientation of the electric vector position angle (E-field)
and are proportional in length to the fractional polarization 
($1''\simeq10$\%).
Bottom: Trend of the fractional polarization as a function of the 
observing wavelengths, at different source locations.
The depolarization $DP$ has been calculated as the ratio of the fractional polarization between the two extreme frequencies.
}
\label{a2065a}
\end{figure*}

The ROSAT HRI X-ray image presented in Fig. \ref{A2065} 
shows the X-ray emission in the $0.1-2.4~ {\mathrm keV}$ band of A2065.
This is one of the galaxy clusters that make up the
Corona Borealis super-cluster.
The cluster X-ray surface brightness of A2065 is consistent with the
optical galaxy surface-density profile, both showing elongation in the
northwest - southeast direction.
The cluster is a centrally condensed system with two dominant
galaxies separated on the plane of the sky by 
17\arcsec~and by $\simeq600~ {\mathrm km/sec}$ in redshift 
(Postman et al. 1988).
An ASCA X-ray observation of A2065 provides evidence for an 
ongoing merger, but also suggests two surface brightness peaks coincident 
with the two central galaxies, apparently survived the merger 
shock passage (Markevitch et al. 1999). 
A Chandra observation (Chatzikos et al. 2006) reveals 
only one X-ray surface brightness peak, which is associated 
with the more luminous, southern galaxy. The gas related to that peak
is cool and displaced slightly from the position of the galaxy. 
We note that the cluster temperature given in Table 1 
(Markevitch 1998) is consistent with results from the ASCA and
Chandra spectrum (White 2000, Ikebe et al.
2002, Chatzikos et al. 2006). On the other hand, David et al. (1993) derived 
a temperature of $kT=8.4^{+3.3}_{-1.8}~ {\mathrm keV}$, from the Einstein MPC detector.

No evidence for the presence of a cluster radio halo is available in the 
literature for A2065.

In this work we have analyzed the radio galaxy labeled with A2065A. It
is located at a projected distance of about 13.6$'$ from the cluster
center. The high resolution image at $6\, {\mathrm cm}$, overlaid on a R-band image
taken with the Telescopio Nazionale Galileo (TNG; Fig. \ref{A2065}), shows a radio galaxy
of about 1$'$ in size. It shows a FRII structure, with a core
located at the position R.A.(J2000)=15h22m46.3s,
Decl.(J2000)=27\degrees55\arcmin40\arcsec 
and two opposite lobes, with evidence of hot-spots at the boundary of the lobes.

This radio galaxy does not show any optical counterpart in the DSS2 red
image. The deeper, 30 min exposure R-band image we have taken with the
TNG shows a faint source with a magnitude R=22.0$\pm$0.3. The small
size of the optical counterpart in comparison with the cluster
galaxies suggests that this object is a cluster background
source. 
Moreover, if we assume that the radio source is at the cluster
redshift, the estimated absolute red magnitude (with no K-correction
applied) would be $M_R\simeq22.0+5-5\times log_{10}D_{L (pc)}=-15.6$. This is a
very faint magnitude. Considering that the typical absolute magnitude
of a FRII is $M_R=-23.6$ (e.g. Scarpa et al. 2002), the apparent
magnitude of this radio source is explained if it is located at $z>1$.

Fig. \ref{a2065a} shows the polarization 
images at 8465 ${\mathrm MHz}$ and 4535 ${\mathrm MHz}$, with an angular resolution of $3''\times3''$.
The polarization has been detected in the lobes at each frequency.
The two lobes have a similar fractional polarization.
In the northern lobe the fractional polarization has a mean value 
of $16\pm5$\% both at 8465 ${\mathrm MHz}$ and 4535 ${\mathrm MHz}$, while in the southern lobe the  fractional polarization has a mean value of $19\pm5$\% at 8465 ${\mathrm MHz}$ and  $17\pm5$\% at 4535 ${\mathrm MHz}$.
The absence of depolarization is also confirmed in the bottom panels 
of Fig. \ref{a2065a} where we show the trend of the fraction polarization 
as a function of the observing frequencies, 
calculated in three different locations of the source.

\subsection{Ophiuchus}

\begin{figure*}
\vspace{9cm}
\includegraphics{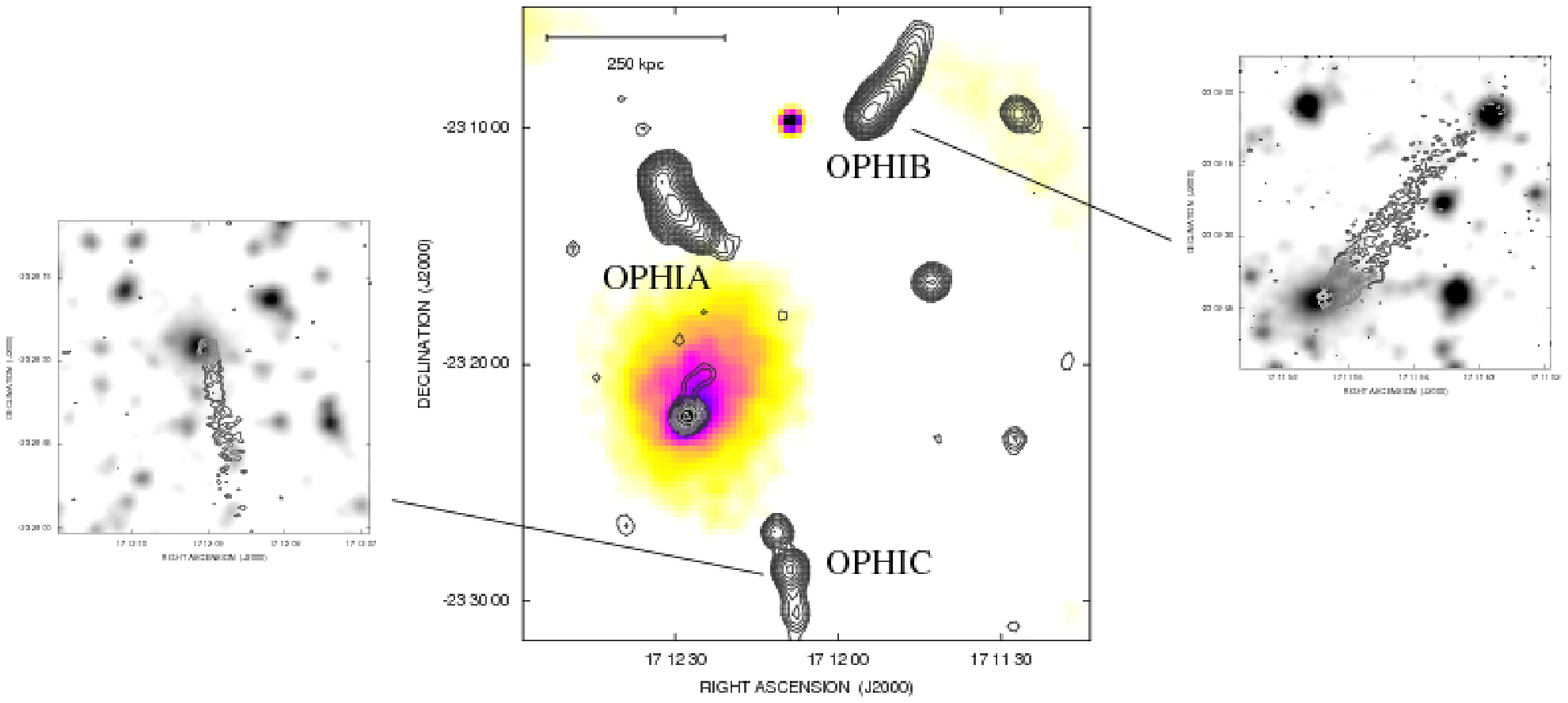}
\caption[]{
1.4 GHz radio image of Ophiuchus obtained from the NVSS (contours) 
overlaid on the ROSAT HRI X-ray image (colors) in the $0.1-2.4~ {\mathrm keV}$ band.
The NVSS image has an angular resolution of 45\arcsec.
The first radio contour is drawn at 1.5 mJy/beam and the rest are spaced by a
factor of $\sqrt 2$. 
The high resolution observations at $6\, {\mathrm cm}$ (contours) overlaid on the DSS2 red plate 
(grey-scale) of the two radio galaxies are inset.
Contour levels = 0.1 0.15 0.3 0.5 0.8 1 mJy/beam.
OPHIB : Angular resolution = $1.09'' \times 0.75''$ (PA=$-$87.0).
Sensitivity = 0.038 mJy/beam. 
OPHIC : Angular resolution = $1.09'' \times 0.76''$ (PA=$-$87.7).
Sensitivity = 0.030 mJy/beam. 
}
\label{OPHI}
\end{figure*}

Ophiuchus is a nearby, hot cluster.
Suzaku data (Fujita et al. 2008b) indicate that this cluster has a cool,
dense core.
We recently found at the cluster center 
a low surface brightness diffuse emission that it is 
classified has a mini-halo (Govoni et al. 2009, Murgia et al. 2009, 
Murgia et al. 2010b). 
Moreover for this cluster there have been 
claims of detecting hard X-ray emission in excess to the thermal emission
(Eckert et al. 2008, Nevalainen et al. 2009). Such emission may be
of non-thermal origin, for example, Compton scattering of relativistic 
electrons by the cosmic microwave background radiation (see e.g.
Rephaeli et al. 2008, Petrosian et al. 2008, and references therein
for recent reviews), but alternative explanations have 
been put forward (Profumo 2008, P{\'e}rez-Torres et al. 2009, Colafrancesco \& Marchegiani 2009).

Fig. \ref{OPHI} shows the ROSAT HRI X-ray image of the cluster
overlaid on the NVSS image.
In this work we have analyzed the radio galaxies labeled with OPHIB and 
OPHIC.
In the original sets of data we observed also OPHIA (see Fig \ref{OPHI}),
but this source is fully resolved out at both wavelengths.
The lack of detection indicates that the OPHIA source is too extended 
to be imaged at a such small angular scale.

The source OPHIB is located to the northwest of the cluster center
at a projected distance of about 14.5\arcmin. 
The high resolution image at $6\, {\mathrm cm}$ overlaid on the DSS2 red plate
shows a source of about 45\arcsec~in size with a NAT head-tail morphology.
The core at the position R.A.(J2000)=17h11m55.4s, 
Decl.(J2000)=$-$23\degrees09\arcmin42.5\arcsec~is coincident with a 
bright galaxy, while the tail is oriented away from the core in the
northwest direction.

The NAT head-tail OPHIC is located
at a projected distance of about 7.6\arcmin~to the southwest of the
cluster center with the head pointed versus the 
cluster center and the tail elongated versus south.
The high resolution image at $6\, {\mathrm cm}$ shows the core, at the position
R.A.(J2000)=17h12m09.0s, Decl.(J2000)=$-$23\degrees28\arcmin26.5\arcsec,
coincident with a galaxy. At $6\, {\mathrm cm}$ the source is about 40$''$ in size. 

\begin{figure*}
\vspace{12cm}
\includegraphics{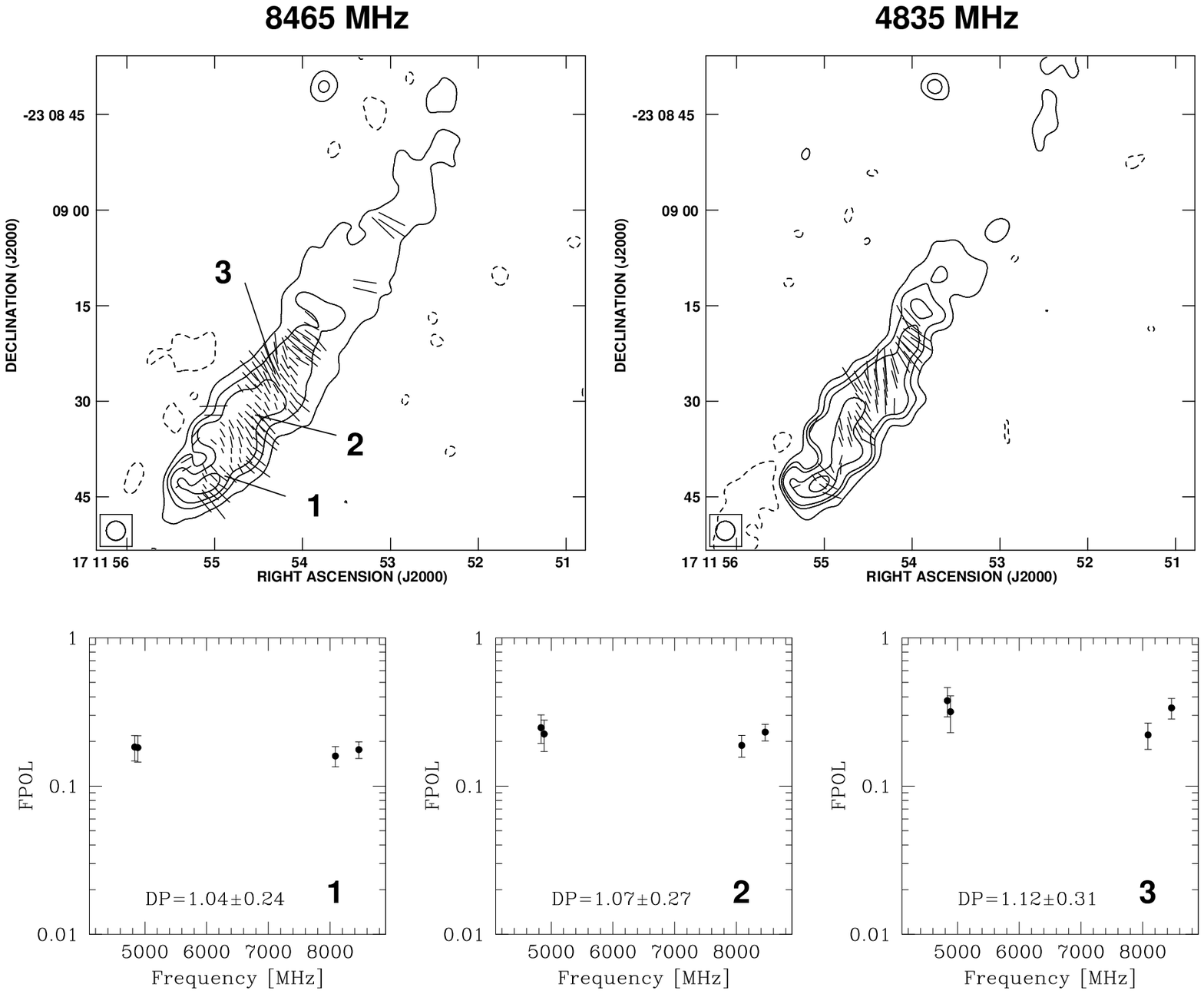}
\caption[]{
Source OPHIB.
Top, left: Total intensity contours and polarization vectors 
at $3.6\, {\mathrm cm}$ (8465 ${\mathrm MHz}$). Contour levels
are drawn at: -0.15 0.15 0.5 1 2 and 3 mJy/beam.
Top, right: Total intensity contours and 
polarization vectors at $6\, {\mathrm cm}$ (4835 ${\mathrm MHz}$). 
Contour levels are drawn at: -0.35 0.35 0.7 1 2 and 3 mJy/beam.
The angular resolution is $3.0''\times 3.0''$. The lines give the
orientation of the electric vector position angle (E-field)
and are proportional in length to the fractional polarization 
($1''\simeq10$\%).
Bottom: Trend of the fractional polarization as a function of the 
observing wavelengths, at different source locations.
The depolarization $DP$ has been calculated as the ratio of the fractional polarization between the two extreme frequencies.
}
\label{ophib}
\end{figure*}

\begin{figure*}
\vspace{8.5cm}
\includegraphics{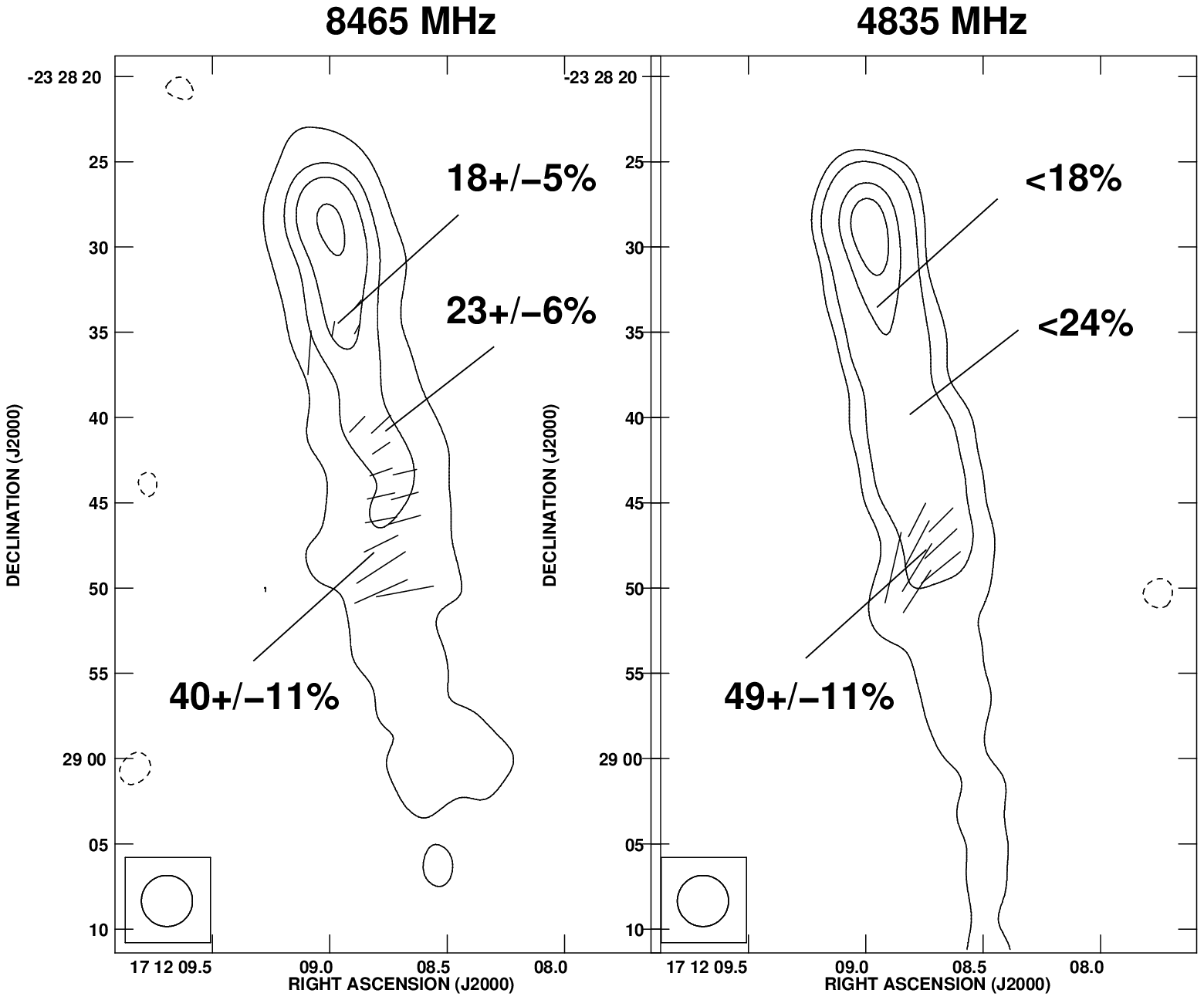}
\caption[]{
Source OPHIC.
Left: Total intensity contours and polarization vectors 
at $3.6\, {\mathrm cm}$ (8465 ${\mathrm MHz}$). Contour levels
are drawn at: -0.1 0.1 0.5 1 and 2 mJy/beam.
Right: Total intensity contours and 
polarization vectors at $6\, {\mathrm cm}$ (4835 ${\mathrm MHz}$). 
Contour levels are drawn at: -0.3 0.3 0.7 2 and 3 mJy/beam.
The angular resolution is $3.0''\times 3.0''$. The lines give the
orientation of the electric vector position angle (E-field)
and are proportional in length to the fractional polarization 
($1''\simeq17$\%).}
\label{ophic}
\end{figure*}

Figs. \ref{ophib} and \ref{ophic} show the polarization 
images at 8465 ${\mathrm MHz}$ and 4835 ${\mathrm MHz}$, with an angular resolution of $3''\times3''$.

OPHIB is polarized at both frequencies up to a distance of about 45$''$.
By considering all the pixels where the total intensity signal
is above 5$\sigma_{\mathrm I}$ at all frequencies, the source has a mean 
fractional polarization of $18\pm5$\% at 8465 
and $22\pm5$\% at 4835 ${\mathrm MHz}$. Within the errors the two
fractional polarizations are compatible therefore in OPHIB we do not detect 
significant depolarization.
In the bottom panels of Fig. \ref{ophib} we show the trend of the fractional 
polarization as a function of the observing frequencies, at different 
source locations.

The fractional polarization in different locations of the source OPHIC 
is indicated in Fig. \ref{ophic}. At 8465 ${\mathrm MHz}$ the polarization is detected only in a few places and the error on the fractional 
polarization is rather high. 
In most of the source the fractional polarization at 4835 is below 
the 3$\sigma_{\mathrm FPOL}$ level. The only patch where a significant 
polarization is detected has a fraction polarization compatible, 
within the errors, with that at 8465 ${\mathrm MHz}$.

\section{Rotation measure images}

\begin{table*}
\caption{Rotation measure results.}
\label{rmtab}
\centering
\begin{tabular} {c r r c r r} 
\hline
Source   & Distance & N     & $<$Errfit$>$  & $\langle RM \rangle$   &  $\sigma_{RM}$         \\
         &  (kpc)   &       & (rad/m$^{2}$) & (rad/m$^{2}$)  &  (rad/m$^{2}$)          \\
\hline  
A401A    & 440      & 34    & 30               & 222$\pm$20     & 113$\pm$14            \\
A401B    & 730      & 23    & 41               & 107$\pm$15       & 74$\pm$14            \\
A2142A   & 270      &  8    & 36               & $-$461$\pm$82    & 230$\pm$63              \\  
A2065A   & 1120     & 12    & 29               & $-9$$\pm$16     &  48$\pm$15              \\ 
OPHIB    & 480      & 15    & 44               & $-$64$\pm$22     &  74$\pm$19               \\  
\hline
\multicolumn{6}{l}{\scriptsize Col. 1: Source; Col. 2: Projected distance from the X-ray center;}\\
\multicolumn{6}{l}{\scriptsize Col. 3: Number of beams over which $RMs$ are computed;}\\ 
\multicolumn{6}{l}{\scriptsize Col. 4: Mean value of the $RM$ fit error;}\\
\multicolumn{6}{l}{\scriptsize Col. 5: Mean of the $RM$ distribution;}\\ 
\multicolumn{6}{l}{\scriptsize Col. 6: RMS of the $RM$ distribution;}\\
\end{tabular}
\end{table*}

The presence of a magnetized and ionized screen between the 
observer and a radio source changes the properties of the incoming 
polarized emission.
As it passes through the magnetized plasma, the polarized synchrotron radiation undergoes the following rotation of the plane of polarization:

\begin{equation}
\Psi _{\mathrm Obs}(\lambda)=\Psi _{\mathrm Int}+\lambda^2 \times RM
\label{psi}
\end{equation}

\noindent
where $\Psi _{\mathrm Obs}(\lambda)$ is the observed polarization angle
at a wavelength $\lambda$ and $\Psi _{\mathrm Int}$ is the 
intrinsic polarization angle. 

The rotation measure $RM$ is related to the plasma thermal electron 
density, $n_{e}$, 
and magnetic field along the line-of-sight, $B_{\|}$, by the equation:
\begin{equation}
RM = 812\int\limits_0^L n_{e} B_{\|} {\mathrm d}l ~~~{\mathrm rad~m}^{-2}
\label{rm}
\end{equation}
where $B_{\|}$ is measured in ${\mathrm \mu G}$, $n_{e}$
in ${\mathrm cm^{-3}}$ and $L$ is the depth of the screen in ${\mathrm kpc}$.   

Following the definition, the $RM$ images were obtained by performing a
fit of the 4 polarization angle images, at each pixel,
as a function of $\lambda^{2}$ (see Eq.\ref{psi}),
by using the software FARADAY by Murgia et al. (2004).
Given as input the images of $Q$ and $U$, at each frequency, the software produces
the $RM$ and the intrinsic polarization angle $\Psi _{\mathrm Int}$ images, both with 
relative error images. To improve the $RM$ image,
the software can be iterated in several self-calibration cycles. In the first cycle
only pixels with the highest signal-to-noise ratio are fitted. In the next cycles 
the algorithm uses the $RM$ information in these high signal-to-noise pixels 
to solve the n$\pi$-ambiguity in adjacent pixels of lower signal-to-noise, in a 
similar method used in the PACERMAN algorithm by Dolag et al. (2005b).

We derived the $RM$ images, at 3\arcsec~resolution,
of the five polarized sources. 
They were calculated only in those pixels in which the following three
conditions were satisfied: the total intensity 
signal at 8465 ${\mathrm MHz}$ was above 3$\sigma_I$, 
the error in the polarization angle at each 
frequency was lower than 15\degrees~, and the resulting $RM$ error
was lower than 60 rad/m$^2$.
We checked the agreement between the final $RM$ images obtained with FARADAY 
and those obtained with the algorithm PACERMAN by Dolag et al. (2005b).

Figs. 11$-$15 show the resulting $RM$ images for the sources A401A, A401B,
A2142A, A2065A and OPHIB, respectively.
For each source the total intensity contours at 8465 ${\mathrm MHz}$
are overlaid on the $RM$ images.
We can characterize the $RM$ distribution in terms of a mean 
($\langle RM \rangle$) and root mean square ($\sigma_{RM}$). 
The histograms of the $RM$ distribution are also shown. In addition, a few plots
showing the position angle  $\Psi _{\mathrm Obs}$ as a function 
of $\lambda^2$ at different source locations are presented.

In Table 4, for each radio galaxy we report its projected distance from the 
cluster X-ray center, the number of beams over which $RMs$ are computed,
the mean value of the $RM$ fit error, the mean ($\langle RM \rangle$) and root 
mean square ($\sigma_{RM}$) of the $RM$ distribution. 
The errors on the $\langle RM \rangle$ and $\sigma_{RM}$ quantities have been computed with Montecarlo simulations, 
considering both the uncertainty related to the presence of the statistical error 
and the fit error\footnote{Indeed, our data suffer from two kinds of uncertainty.
The first source of error is related to the fit
of the $\lambda^2$-law and takes into account
the presence of errors in the measurements of $Q$ and $U$.
The fit error has
the effect to broad the histogram of the $RM$ distribution and thus to
increase the observed $\sigma_{RM}$. While the second uncertainty
is the statistical error. The statistical errors for $\langle RM \rangle$ and for $\sigma_{RM}$ 
is given by $\sigma_{RM}/ \sqrt{N}$ and $\sigma_{RM}/ \sqrt{2N}$ respectively, where
$N$ is the number of beams over which the RM has been computed.
Through the Montecarlo procedure we 
checked that the fit error is negligible with respect to the statistical error.}.

The values of $\langle{\mathrm RM}\rangle$ given in Table 4 are not corrected for the contribution of the Galaxy. 
We determined the Galactic contribution for each cluster of our sample according to the values reported
in the $RM$ catalog by Taylor et al. (2009). We used the distance-weighted mean
of the $RMs$ of all sources located within a radius of $3\deg$.
The $RM$ Galactic contribution in the region occupied by our cluster sample has been found to be rather small.
About $-6$ rad/m$^2$ for A401, $+16$ rad/m$^2$  for A2142, 
and $+10$ rad/m$^2$ for A2065. While the Galactic RM
is $-29$ rad/m$^2$ for Ophiuchus.

In order to verify the polarization angle linearity with $\lambda^2$,
we selected some pixels in different locations of the sources.
The data are generally quite well represented by a linear $\lambda^2$
relation. The polarization angle trend with $\lambda^2$ and 
the absence of depolarization 
(at least in the case of A401, A2065, 
and Ophiuchus) would argue in favor of an RM produced by an external 
Faraday screen (Burn 1966, Laing 1984, Laing et al. 2008). 
However, some authors have suggested the possibility that 
the $RM$ observed in radio galaxies is not associated with the 
foreground intracluster medium, but may arise locally to the radio
source (Bicknell et al. 1990, Rudnick \& Blundell 2003).
A much larger $\lambda^2$ range would be needed 
to unambiguously exclude the presence of an internal Faraday screen.
However, although in this work we cannot address the origin 
of the screen properly (internal, external, or both), 
2D Montecarlo simulations presented in the Appendix show that both $RM$ 
and polarization data can be explained by the presence of a foreground Faraday screen only.
In agreement with the 2D Montecarlo simulations, in the following we assume that the internal contribution is negligible
and that the major factor responsible for the Faraday rotation is given by the external intracluster medium (e.g.
Govoni \& Feretti and references therein).
In this case, the structure of the $RM$ images and the values of the $\sigma_{RM}$
and $\langle RM \rangle$ can give information on the intracluster magnetic field structure.
In particular, the $RM$ dispersion of the histograms and the
presence of small scales structure in the $RM$ images can be explained by the fact
that the cluster magnetic field fluctuates on scales
smaller than the size of the sources.  On the other hand,
the cases of $RM$ distributions (once corrected for the
Galactic contribution) with a non-zero mean indicate that the magnetic 
field fluctuates also on scales larger than the radio sources.
Therefore, in order to interpret correctly
the $RM$ data, magnetic field
fluctuations in a wide range of spatial scales should be considered.

The radio galaxies analyzed in this work are located
at a projected distance from the cluster center which 
range from $270~ {\mathrm kpc}$ (A2142A) up to $1120~ {\mathrm kpc}$ (A2065A).

The case of A401 is very interesting because it allows the $RM$ to 
be sampled along
two different lines-of-sight within the same cluster.
As expected in the case of the $RM$ being due to an external Faraday rotation
due to the cluster magneto-ionized medium, the innermost source (A401A) has a
higher $\sigma_{RM}$ and \absrmm~ than the external one (A401B).
The $RM$ images reveal patchy structures with $RM$ fluctuations
down to scales of a few kpc. 

These small scale $RM$ structures are well 
visible also in the other radio galaxies analyzed here.
In A401A, A401B and A2142 
large patches of fairly constant $RM$ are also visible, indicating a magnetic 
structure fluctuating on large scales.
We note however that the sampling of large scale fluctuations 
is rather poor.

We note that the location of Ophiuchus near the Galactic center makes this target not optimal for a magnetic field structure investigation through the $RM$. Therefore, the $RM$ image of OPHIB should be considered with caution. 

A2065A is a nice example of a high resolution $RM$ image in an extended background source. 
Despite its distance from the cluster center, its importance
is due to fact that, at least in this case, we can exclude that the  patchy structures seen in the $RM$ image
are due to the turbulence locally created by the motion of the galaxy in the intracluster 
medium.

\begin{figure*}
\vspace{20cm}
\includegraphics{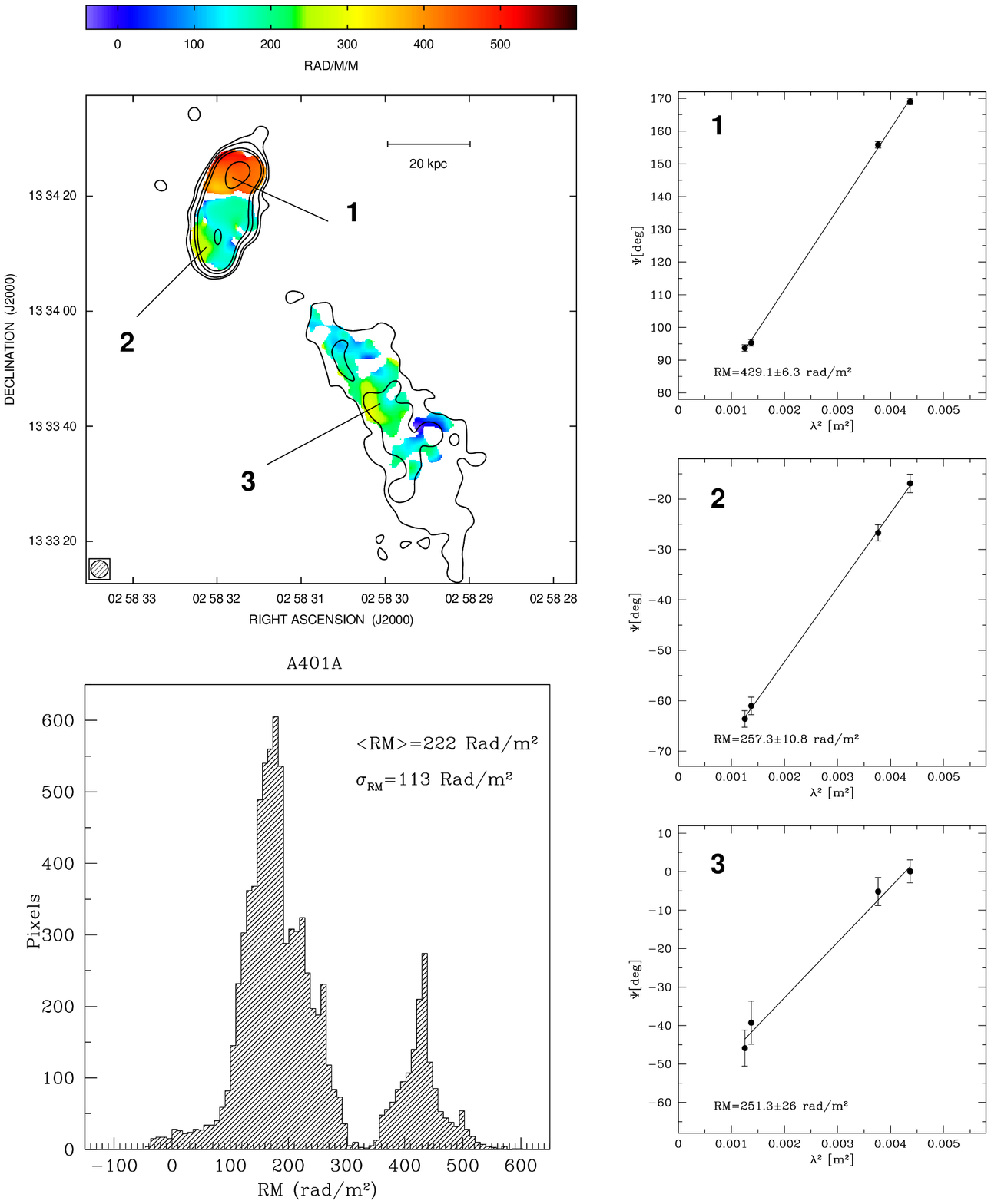}
\caption[]{
Rotation measure image of the radio galaxy A401A.
The angular resolution is $3.0''\times 3.0''$. 
The contours refer to the total intensity image at 8465 ${\mathrm MHz}$.
Contour levels are drawn at: 0.07 0.15 0.5 1 and 5 mJy/beam.
The histogram shows the rotation measure distribution for 
all significant pixels.
The plots show the position angle  
$\Psi _{\mathrm Obs}$ as a function of $\lambda^2$ at different source locations.
}
\label{a401A}
\end{figure*}

\begin{figure*}
\vspace{20cm}
\includegraphics{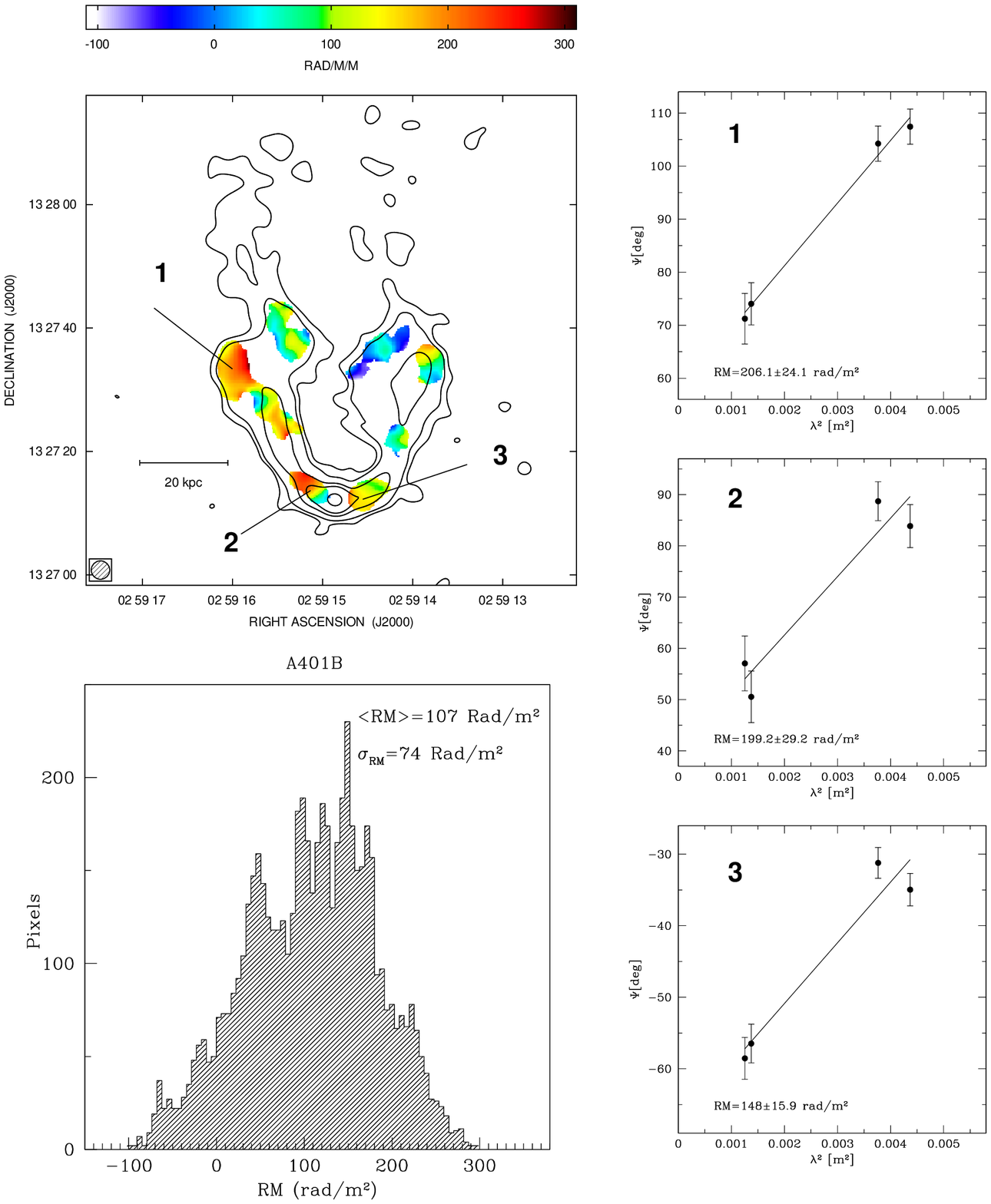}
\caption[]{
Rotation measure image of the radio galaxy A401B.
The angular resolution is $3.0''\times 3.0''$. 
The contours refer to the total intensity image at 8465 ${\mathrm MHz}$.
Contour levels are drawn at: 0.07 0.15 0.5 1 and 3 mJy/beam.
The histogram shows the rotation measure distribution for all significant pixels.
The plots show the position angle  
$\Psi _{\mathrm Obs}$ as a function of $\lambda^2$ at different source locations.}
\label{a401B}
\end{figure*}

\begin{figure*}
\vspace{20cm}
\includegraphics{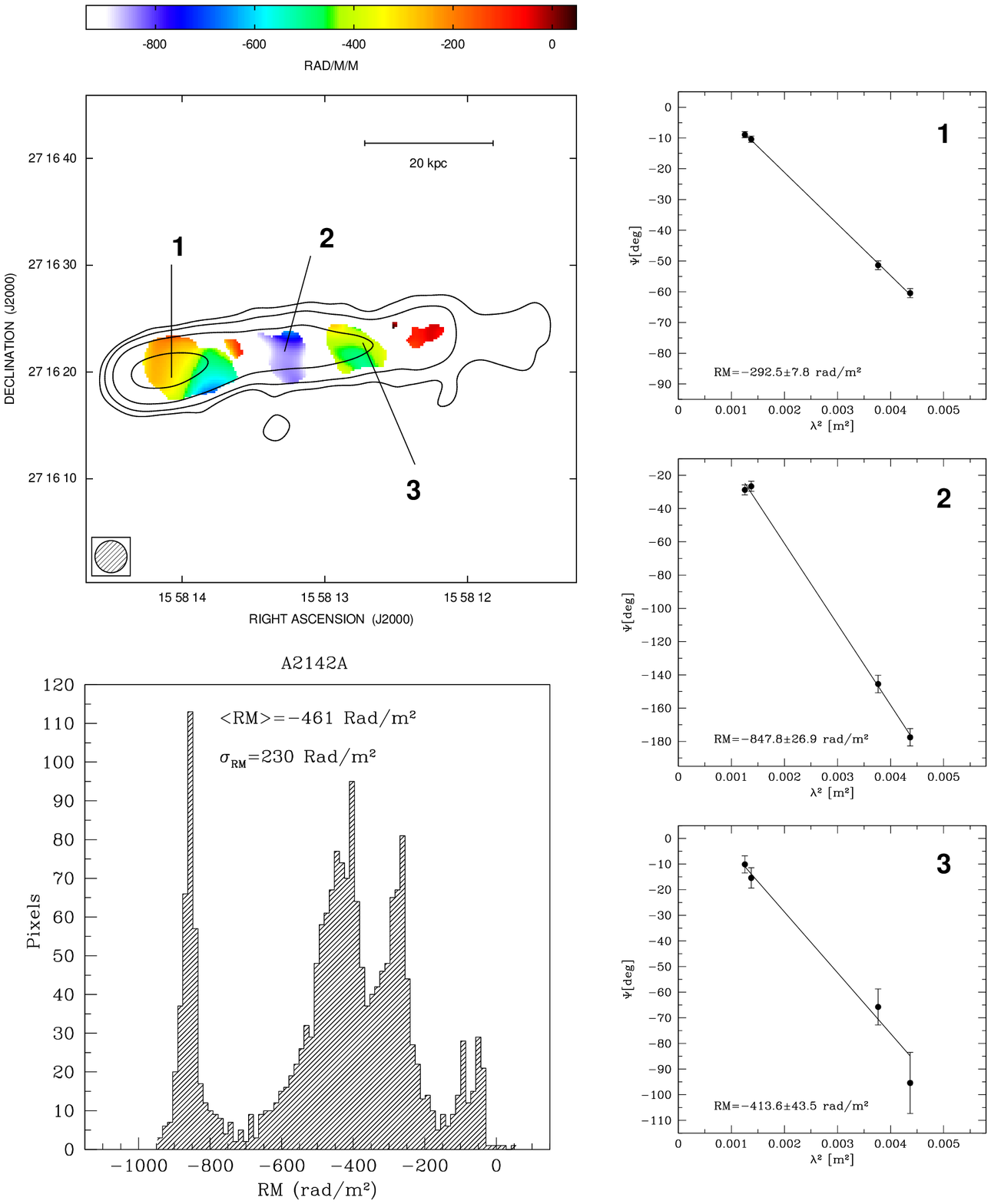}
\caption[]{
Rotation measure image of the radio galaxy A2142A.
The angular resolution is $3.0''\times 3.0''$. 
The contours refer to the total intensity image at 8465 ${\mathrm MHz}$.
Contour levels are drawn at: 0.06 0.15 0.5 1 and 3 mJy/beam.
The histogram shows the rotation measure distribution for all significant pixels.
The plots show the position angle  
$\Psi _{\mathrm Obs}$ as a function of $\lambda^2$ at different source locations.}
\label{a2142A}
\end{figure*}

\begin{figure*}
\vspace{20cm}
\includegraphics{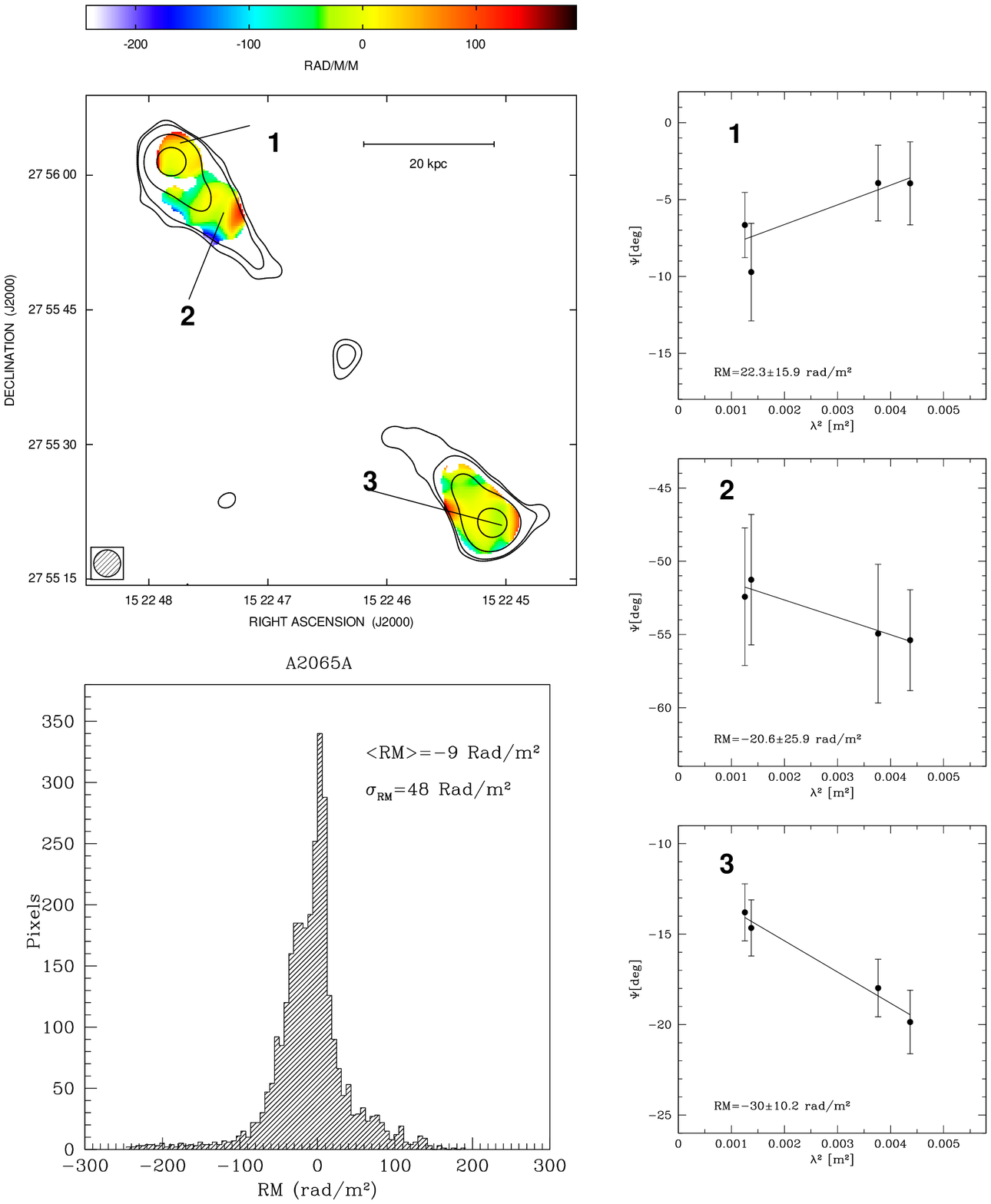}
\caption[]{
Rotation measure image of the radio galaxy A2065A.
The angular resolution is $3.0''\times 3.0''$. 
The contours refer to the total intensity image at 8465 ${\mathrm MHz}$.
Contour levels are drawn at: 0.06 0.1 0.5 and 3 mJy/beam.
The histogram shows the rotation measure distribution for all 
significant pixels.
The plots show the position angle  
$\Psi _{\mathrm Obs}$ as a function of $\lambda^2$ at different source locations.}
\label{a2065A}
\end{figure*}

\begin{figure*}
\vspace{20cm}
\includegraphics{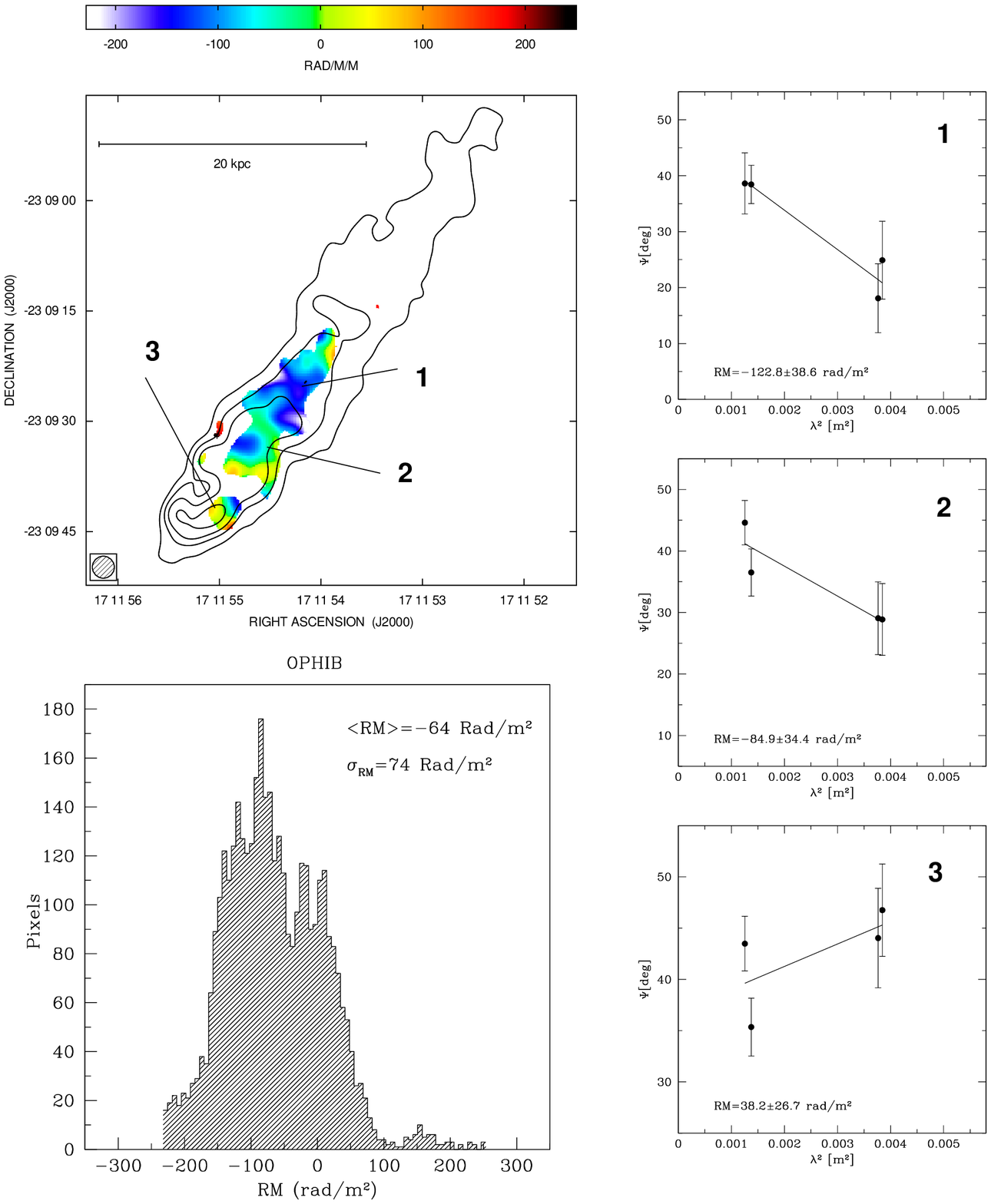}
\caption[]{
Rotation measure image of the radio galaxy OPHIB.
The angular resolution is $3.0''\times 3.0''$.
The contours refer to the total intensity image at 8465 ${\mathrm MHz}$.
Contour levels are drawn at: 0.15 0.5 1 2 and 3 mJy/beam.
The histogram shows the rotation measure distribution for all significant 
pixels. The plots show the position angle  
$\Psi _{\mathrm Obs}$ as a function of $\lambda^2$ at different source locations.}
\label{OPHIB}
\end{figure*}

\section{Magnetic field - gas temperature connection}

The new $RM$ data presented here increase the statistic of the $RM$ for radio sources in hot clusters. 
Therefore, by comparing these new data with $RM$ information taken from the
literature it is now possible to investigate the connection between 
the magnetic field strength and the temperature of the intracluster medium.
For a detailed statistical analysis, we formed a sample of 12 galaxy clusters for which high-quality $RM$ data are available.

The cluster X-ray properties of the sample are presented in Table 5.
For the distribution of the thermal electron gas density we assumed 
the standard 
$\beta$-model profile (Cavaliere \& Fusco-Femiano 1976): 
\begin{equation}
n_e(r)=n_0(1+r^2/r_c^2)^{-3\beta/2}
\end{equation}
where $n_0$ and $r_c$ are the central electron density, and
the cluster core radius, respectively. The X-ray parameters reported in Table 5 are taken from the literature and corrected for the cosmology adopted in this paper. The mean cluster density calculated by integrating the 
$\beta$-model profile over a sphere of 1Mpc in radius is also indicated.
 
In the sample the temperature of the clusters is in the range 
$T\simeq3-10$ keV, while the mean $\beta$-model parameters are:
$<\beta>$=0.7, $<r_c>$=285 kpc, and $<n_0>=4.4\times 10^{-3}$ cm$^{-3}$.
In the cluster sample there are a total of 29 radio galaxies, for which high-quality $RM$ data are available.
For each radio source, in Table 6 we report the projected distance from the cluster X-ray center, 
and the $\sigma_{RM}$ taken from the literature.
We included in this analysis only those clusters
hosting radio sources for which high-resolution, extended $RM$ images are
available. We did not consider $RM$ images of sources located at the center of 
cooling core clusters (e.g. Taylor et al. 2002).
Indeed in our list there are a few cooling flows clusters (A2142, Ophiuchus, A2382),
but in these cases the analyzed sources are located at a large projected distances from the cluster center,
therefore the contribution of the dense cool core to the $RM$ is likely to be negligible.

\begin{figure*}[t]
\centering
\includegraphics[width=15 cm]{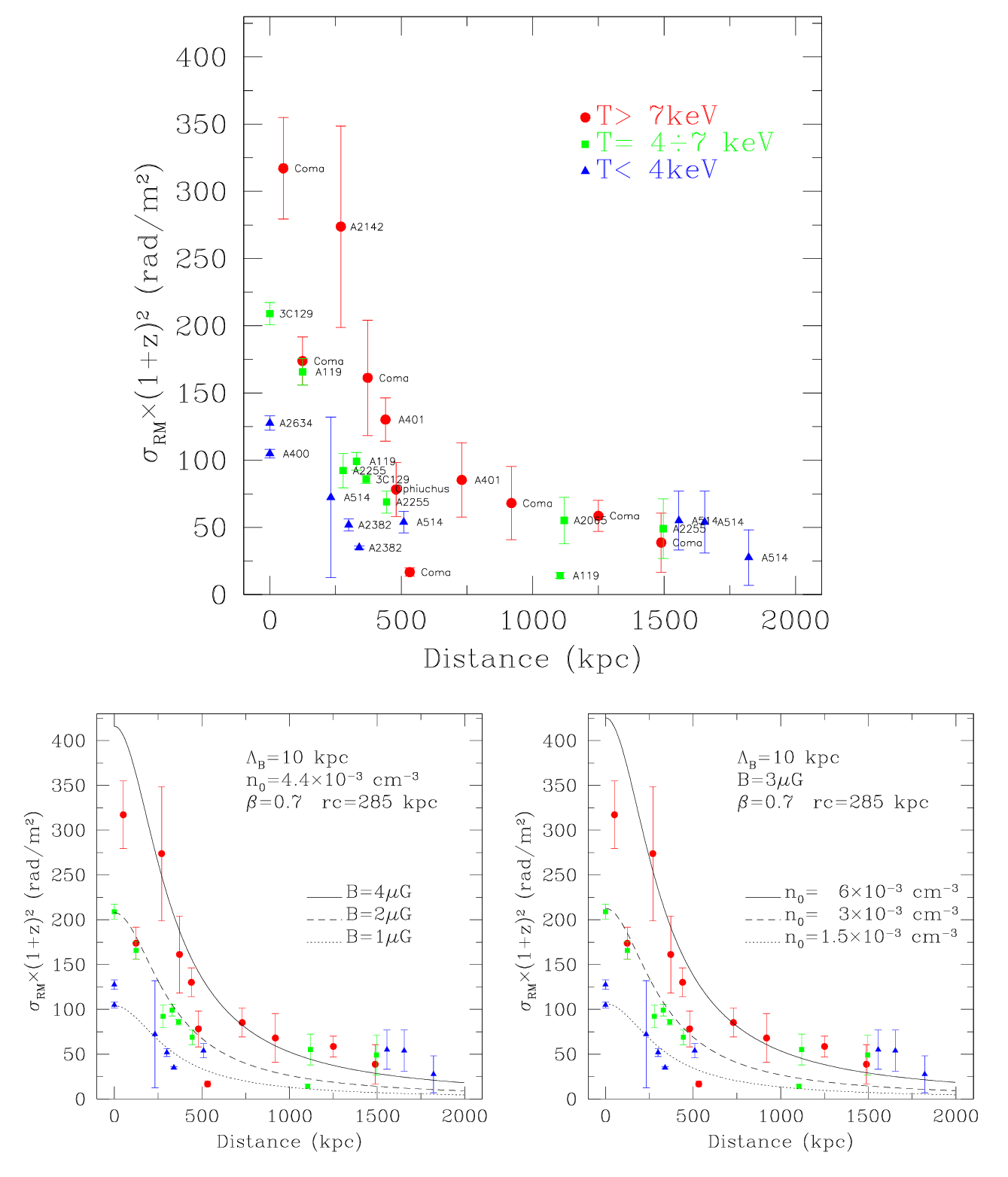}
\caption{Top: Dispersion of the rotation measure distribution 
as a function of the projected distance from the cluster X-ray center. 
The different symbols represent the cluster temperature taken from the 
literature (red $>$ 7 keV, green $4-7$ keV, blue $<$ 4 keV).
Bottom:  Prediction of the analytical formulation by assuming a magnetic field scale 
$\Lambda_{B}=10~ {\mathrm kpc}$, and by keeping fixed the X-ray parameters to the mean values of the cluster 
sample $\beta=0.7$ $r_c=285~ {\mathrm kpc}$. 
On the left $n_0=4.4 \times 10^{-3}~ {\mathrm cm^{-3}}$ has been fixed and
we present the expectations for three different magnetic field strengths ($B=1,2,4~{\mathrm \mu G}$). 
On the right $B=3~ {\mathrm \mu G}$ has been fixed and we present the expectations for three different 
central gas densities ($n_0=1.5,3,6\times10^{-3}~ {\mathrm cm^{-3}}$).}
\label{sigmarm}
\end{figure*}

In Fig.  \ref{sigmarm} (top) we analyze the trend 
of $\sigma_{RM}$, redshift corrected by multiplying by a factor of $(1+z)^2$,
as a function of distance from the cluster X-ray center.
The different symbols represent different cluster temperature.  
It is evident a clear broadening of the $\sigma_{RM}$ toward small projected 
distances, consistently with an excess of Faraday rotation due to the magnetized 
intracluster medium.
In addition, a positive trend is found between the cluster temperature 
and the $\sigma_{RM}$. In particular, for a fixed projected distance from the cluster center, clusters with higher
temperature show a higher $\sigma_{RM}$.

The interpretation of $\sigma_{RM}$ in a cluster depends on several assumptions, including
the model for the X-ray emitting gas and the magnetic field structure.
To try to explain the trend between the cluster temperature 
and the $\sigma_{RM}$ seen in Fig. \ref{sigmarm} (top), for simplicity,
 we can consider an analytical formulation based 
on the approximation that the magnetic field is constant along the cluster
and tangled on a single 
scale $\Lambda_{B}$. 
In this simple ideal case, the screen is made of cells 
of uniform size, electron density and magnetic field strength, but with a field orientation at random
angles in each cell. The observed $RM$ along any given line of sight is
then generated by a random walk process involving a large
number of cells of size $\Lambda_{B}$.  The distribution of the $RM$
is Gaussian with zero mean, and a variance given by:
\begin{equation}
 \sigma_{RM}^{2}= \langle {RM^{2}} \rangle = 812^{2} \Lambda_{B} \int\limits_0^L ( n_{e} B_{\|})^{2} {\mathrm d}l~.
\label{sigma}
\end{equation}
In this formulation, by considering a density distribution which follows a $\beta$-profile,
the following relation (e.g. Lawler \& Dennison 1982, Tribble
1991, Feretti et al. 1995, Felten 1996) for the $RM$ dispersion is 
obtained by integrating Eq.~\ref{sigma}:

\begin{equation} 
\sigma_{RM}(r)= {{K B n_{0}  r_c^{1/2} \Lambda_{B}^{1/2} }
 \over {(1+r^2/r_c^2)^{(6\beta -1)/4}}} \sqrt {{\Gamma(3\beta-0.5)}\over{\Gamma(3\beta)}}
\label{felten}
\end{equation}
where $B=\sqrt 3 B_{\|}$ and $\Gamma$ is the Gamma function. 
The constant $K$ depends on the integration path over 
the gas density distribution:
$K$ = 624, if the source lies completely
beyond the cluster, and $K$ = 441 if the source is halfway the cluster.
Therefore, $\sigma_{RM}$ depends on the 
possible fluctuations in $n_{0}$, $B$, $\sqrt r_c$ and $\sqrt \Lambda_{B}$.
Since the linear dependence on $n_{0}$ and $B$ strongly affect
$\sigma_{RM}$, in the following we concentrate on these two parameters.
For most clusters of our sample, measurements of the magnetic
field scale are not available from the literature.
However, a discussion regarding the effect 
of $\Lambda_{B}$ on $\sigma_{RM}$ is reported in the Appendix.

The trend seen in the top of 
Fig. \ref{sigmarm}, can be explained 
by a gas temperature $-$ gas density relation, 
or by a gas temperature $-$ magnetic field strength relation.

In the bottom panels of Fig. \ref{sigmarm} we show the predictions of the 
analytical formulation (Eq. \ref{felten}) obtained by assuming a magnetic field 
scale $\Lambda_{B}=10~ {\mathrm kpc}$, and by keeping fixed the X-ray parameters 
to the mean values of the cluster sample ($\beta$=0.7, $r_c=285~ {\mathrm kpc}$).
On the left bottom panel, we fixed the central gas density 
to $n_0=4.4 \times10^{-3}~ {\mathrm cm^{-3}}$ (the mean value of the cluster sample) and
we present the expectations for three different magnetic field 
strengths ($B=1,2,4~ {\mathrm \mu G}$).
On the right bottom panel, we fixed $B=3~ {\mathrm \mu G}$ and
we present the expectations for three different central 
gas densities ($n_0=1.5,3,6\times10^{-3}~ {\mathrm cm^{-3}}$).
In both cases the statistical trend of the data can be reproduced quite well.
Therefore, the higher $\sigma_{RM}$ in hotter clusters may be explained if they have 
a higher magnetic field strength and/or if they have a 
higher gas density.

\begin{figure*}
\centering
\includegraphics[width=15 cm]{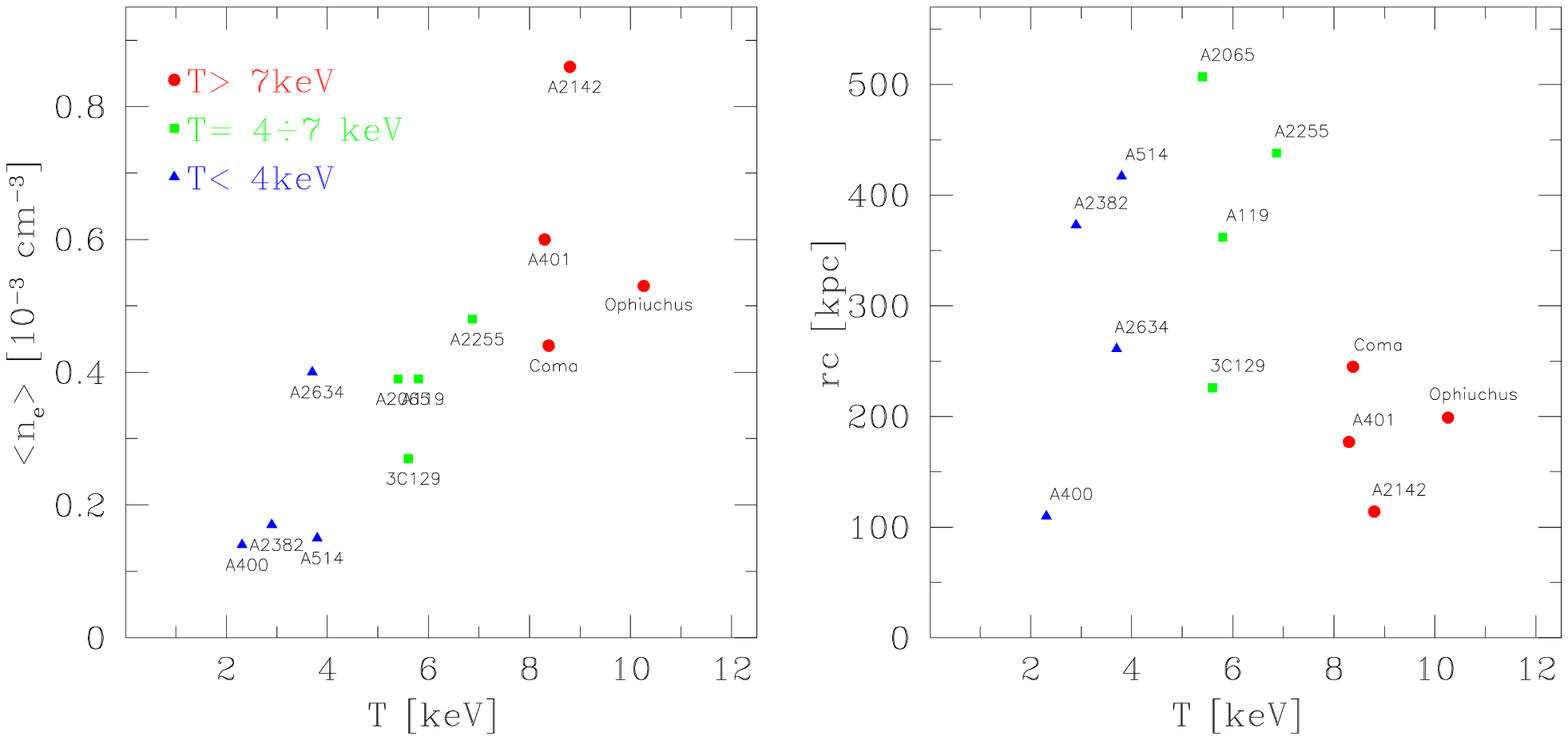}
\caption{
Left: Plot of the mean gas density versus the temperature of the 
clusters in the sample.
The mean cluster density has been calculated by integrating the 
$\beta$-model profile over a sphere of 1Mpc in radius.
Right: Plot of the core radius versus the temperature of the 
clusters in the sample.
The different symbols represent the cluster temperature taken from the 
literature (red $>$ 7 keV, green $4-7$ keV, blue $<$ 4 keV).}
\label{density}
\end{figure*}

It is not easy to disentangle the two effects; however, in the
following we show that hotter clusters have 
higher $\sigma_{RM}$ mostly because of their higher gas density. 
In Fig. \ref{density} (left), we show the mean gas density 
plotted versus the temperature of the clusters.
Clusters with a higher 
temperature show a higher gas density.
The difference of about a factor of four in the mean density between cooler 
and hotter clusters may explain the difference of about a factor of four 
in $\sigma_{RM}$ between cooler and hotter clusters.
Therefore, this analysis does not confirm a strict link between 
the magnetic field strength and the gas temperature 
of the intracluster medium.

In Fig. \ref{density} (right), we show 
the core radius plotted versus the temperature 
of the clusters. No obvious trend is present between
the temperature and core radius of the clusters.

An alternative method to investigate a possible connection between the magnetic field strength and the gas temperature of the intracluster medium 
is the analysis of the $\sigma_{RM}-S_{X}$ correlation.
The cluster X-ray surface brightness $S_{X}$ is given by:
\begin{equation}
S_{X} \propto \int_{LOS} n_{e}^2 \sqrt{\mathrm T}~{\mathrm d}l
\label{sx}
\end{equation}
\noindent

The X-ray surface brightness (see Table 6) has been calculated by analyzing
pointed ROSAT PSPC observations in the $0.1-2.4~ {\mathrm keV}$ band. 
ROSAT PSPC pointed observations were not available in the case of 3C129 and A2065 
therefore we used the images taken from the ROSAT All Sky Survey,
in the same band.
The X-ray surface brightness has not been calculated in the location of the 
sources 5C4.127 and 5C4.42 (in the Coma cluster) because 
they are located outside the field of view of the ROSAT detector. 
The X-ray surface brightness (in Cts s$^{-1}$pixel$^{-2}$) has been derived 
by averaging the X-ray count rates in an annulus comprising the position of the 
radio galaxies. 
We then corrected the count rates for the background. The  background has been 
measured well outside the X-ray-bright cluster region.
X-ray point sources have been masked out in this analysis.
We finally converted the X-ray surface brightness from Cts s$^{-1}$pixel$^{-2}$ to
erg s$^{-1}$cm$^{-2}$sterad$^{-1}$ by using the 
PIMMS software\footnote{http://heasarc.gsfc.nasa.gov/Tools/w3pimms.html}.
In this conversion the X-ray emission was approximated by a Raymond-Smith model
with the mean cluster temperature and photoelectric absorption column density 
given in the literature (see Table \ref{rmsample}).

\begin{figure*}[t]
\centering
\includegraphics[width=12 cm]{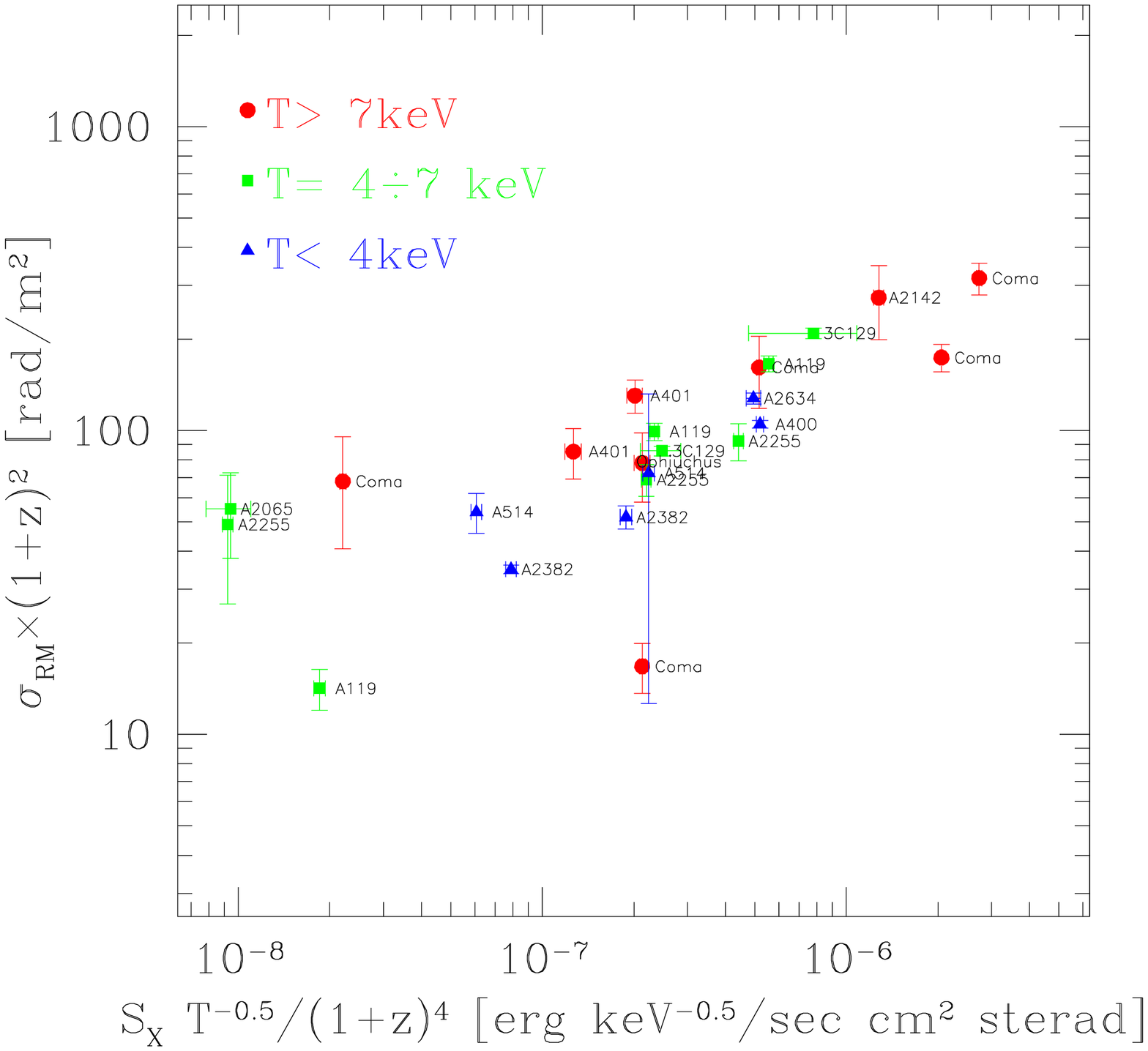}
\caption{Dispersion of the rotation measure distribution 
as a function of the X-ray surface brightness of the intracluster gas in the source location.
The different symbols represent the cluster temperature taken from the 
literature (red $>$ 7 keV, green $4-7$ keV, blue $<$ 4 keV).}
\label{sigmasx}
\end{figure*}

In Fig. \ref{sigmasx} we report the $\sigma_{RM}-S_{X}$ plot
showing the trend of $\sigma_{RM}$ values as a function of the
cluster X-ray surface brightness $S_{X}$ calculated in the
corresponding source location. In the plot, $S_{X}$ has been corrected
for the dependence on the temperature by multiplying by a factor of $T^{-0.5}$.
In addition, $S_{X}$ has been corrected for the
cosmological dimming of the surface brightness by dividing by a factor of $(1+z)^4$.
The new $RM$ data presented  here (A401A, A401B, OPHIB, A2142A, A2065A) confirm the  $\sigma_{RM}-S_{X}$
relation previously known in the literature (e.g. Dolag et al. 2001, Weratschnig et al. 2008).
The different symbols represent different cluster temperature.  
As predicted by cosmological magneto-hydrodynamic simulations (Dolag et al. 1999), 
galaxy clusters may have different central magnetic field strengths
depending on their dynamical state, leading to
an offset  of the $\sigma_{RM}-S_{X}$ relation, along the $\sigma_{RM}$ axis, depending on the cluster temperature.
An offset of the $\sigma_{RM}-S_{X}$ relation, along the $\sigma_{RM}$ axis, depending on the cluster 
temperature if present is comparable to the intrinsic scatter of the correlation. Therefore, a possible connection between the magnetic field strength and the gas temperature, 
if present, is very weak. 
In addition, although the statistics is still limited, we note
that in the $\sigma_{RM}-S_{X}$ relation
clusters with extended diffuse emission (e.g. Coma, A2255, A401)
seem to show a behavior similar to that of
clusters without diffuse synchrotron emission.
   
In conclusion, present data does not permit to establish
a strict link between the 
magnetic field strength and the gas temperature of the intracluster medium. 

\begin{table*}
\caption{X-ray data of the cluster sample.}
\label{rmsample}
\centering
\begin{tabular} {llcccccccc} 
\hline
 Cluster  &     z    &kpc/\arcsec& T      &  nH    & $\beta$ & r$_c$ & n$_0$               &  $<n_e>$         & Reference    \\
          &          &           & [keV]  &        &         & kpc   & 10$^{-3}$ cm$^{-3}$ &  10$^{-3}$ cm$^{-3}$ &              \\
\hline
A514      &   0.0714 & 1.34   & 3.8     & 3.64$\times$10$^{20}$    & 0.6    & 417  &  0.5 & 0.15  & 1,7 \\
Coma      &   0.0232 & 0.46   & 8.38    & 8.54$\times$10$^{19}$    & 0.654  & 245  &  3.5 & 0.44  & 2,8  \\
A2255     &   0.0806 & 1.50   & 6.87    & 2.49$\times$10$^{20}$    & 0.797  & 438  &  2.1 & 0.48  & 3,8 \\
A400      &   0.0240 & 0.48   & 2.31    & 8.33$\times$10$^{20}$    & 0.534  & 110  &  2.4 & 0.14  & 2,8\\
A2634     &   0.0312 & 0.61   & 3.7     & 4.79$\times$10$^{20}$    & 0.640  & 261  &  2.8 & 0.40  & 2,8 \\
A119      &   0.0441 & 0.86   & 5.8     & 3.51$\times$10$^{20}$    & 0.675  & 362  &  1.8 & 0.39  & 4,8 \\
3C129     &   0.0223 & 0.44   & 5.6     & 5.96$\times$10$^{21}$    & 0.601  & 226  &  2.1 & 0.27  & 5,8 \\
A2382     &   0.0618 & 1.18   & 2.9     & 3.99$\times$10$^{20}$    & 0.9    & 373  &  1.2 & 0.17  & 6,9 \\
A401      &   0.074  & 1.39   & 8.3     & 9.88$\times$10$^{20}$    & 0.613  & 177  &  7.0 & 0.60  & 4,8  \\
A2142     &   0.091  & 1.67   & 8.8     & 3.78$\times$10$^{20}$    & 0.591  & 114  & 18.7 & 0.86  & 4,8  \\
A2065     &   0.073  & 1.37   & 5.4     & 3.04$\times$10$^{20}$    & 1.162  & 507  &  2.3 & 0.39  & 4,8  \\
Ophiuchus &   0.028  & 0.55   & 10.26   & 1.93$\times$10$^{21}$    & 0.747  & 199  &  8.0 & 0.53  & 2,8  \\   
\hline
\multicolumn{10}{l}{\scriptsize Col. 1: Cluster name; Col. 2: Redshift;  Col. 3: Angular to linear conversion; Col. 4: Cluster temperature; }\\
\multicolumn{10}{l}{\scriptsize Col. 5: Galactic absorption taken from the Leiden/Argentine/Bonn (LAB) Survey of Galactic HI (Kalberla et al. 2005);}\\ 
\multicolumn{10}{l}{\scriptsize Col. 6, 7:  $\beta$-model parameters ($\beta$ and core radius); Col. 8: Central gas density;}\\  
\multicolumn{10}{l}{\scriptsize Col. 9: Mean gas density calculated by integrating the $\beta$-model profile over a sphere of 1Mpc in radius;}\\
\multicolumn{10}{l}{\scriptsize Col. 10: Temperature and $\beta$-model references:}\\
\multicolumn{10}{l}{\scriptsize 1 Weratschnig et al. (2008), 2 Fukazawa et al. (1998), 3  White (2000), 4  Markevitch (1998),}\\ 
\multicolumn{10}{l}{\scriptsize 5 Edge \& Stewart (1991), 6 Ebeling et al. (1996),}\\
\multicolumn{10}{l}{\scriptsize 7 Govoni et al. (2001), 8 Chen et al. 2007, 9 Guidetti et al. (2008)}\\
\end{tabular}
\end{table*}

\begin{table*}
\caption{$RM$ and X-ray data.}
\centering
\begin{tabular} {llrrcc} 
\hline
 Cluster  &Source  &  Distance &   $\sigma_{RM}$ &  $S_X$  [$0.1-2.4$] keV       & Reference    \\
          &        &  [kpc]    &  [rad/m$^{2}$]  &  $10^{-7}$[erg/ s cm$^2$ sterad] &              \\
\hline
A514      & A514B2          &   232     &   63   &  5.75$\pm$0.13    & 1 \\
          &  A514D          &   509     &   47   &  1.56$\pm$0.03    & 1    \\ 
          & A514A           &  1556     &   48   &  $<$1.79          & 1   \\
          & A514E           &  1654     &   47   &  $<$1.79          & 1   \\ 
          &  A514C          &  1822     &   24   &  $<$1.79          & 1   \\
Coma      & 5C4.85          &   51    &    303  &  86.74$\pm$0.48     & 2  \\  
          &  5C4.81,NGC4869 &  124    &    166  &  65.22$\pm$0.26     & 2      \\
          &  5C4.74         &  372    &    154  &  16.37$\pm$0.08     & 2      \\
          &  5C4.114        &  532    &     16  &   6.76$\pm$0.05     & 2      \\
          &  5C4.127        &  919    &     65  &   0.70$\pm$0.01    & 2      \\
          &  5C4.42         &  1250   &     56  &     -              & 2      \\
          &   5C4.152       &  1489   &     37  &     -              & 2      \\
A2255     & 1712.4+6401     &   279     &   79   &  15.78$\pm$0.22   & 3 \\
          & J1713.5+6402    &   444     &   59   &   7.86$\pm$0.11   & 3     \\
          &  J1713.3+6347   &  1497     &   42   &   0.33$\pm$0.01   & 3     \\
A400      & 3C75            &     0     &   100  &   8.69$\pm$0.15   & 4\\
A2634     & 3C465           &     0     &   120  &  10.77$\pm$0.30   & 4 \\
A119      & 0053-015        &   124     &  152   &  15.89$\pm$0.22   & 5 \\
          &  0053-016       &   330     &   91   &   6.68$\pm$0.08   & 5     \\
          &  3C29           &  1104     &   13   &   0.53$\pm$0.01   & 5     \\
3C129     & 3C129.1         &     0     &   200  &  20.14$\pm$3.31   & 6 \\
          &  3C129          &   367     &   82   &   6.40$\pm$0.41   & 6     \\
A2382     & PKS2149-158C    &   300     &   46   &   4.08$\pm$0.10   & 7 \\
          & PKS2149-158     &   340     &   31   &   1.71$\pm$0.04   & 7     \\
A401      & A401A           &   440     &  113   &   7.72$\pm$0.16   & *  \\
          &  A401B          &   730     &   74   &   4.84$\pm$0.10   & *     \\
A2142     & A2142A          &   270     &  230   &  53.72$\pm$0.69   & *  \\
A2065     & A2065A          &  1120     &   48   &   0.29$\pm$0.02   & *  \\
Ophiuchus & OPHIB           &   480     &   74   &   7.63$\pm$0.14   & *  \\   
\hline
\multicolumn{6}{l}{\scriptsize Col. 1: Cluster name; Col. 2: Source name/label;}\\
\multicolumn{6}{l}{\scriptsize Col. 3: Projected distance from the cluster X-ray center;}\\ 
\multicolumn{6}{l}{\scriptsize Col. 4: RMS of the $RM$ distribution;}\\ 
\multicolumn{6}{l}{\scriptsize Col. 5: Cluster X-ray surface brightness of the intracluster gas in the source location;}\\
\multicolumn{6}{l}{\scriptsize Col. 6: $\sigma_{RM}$  reference:}\\
\multicolumn{6}{l}{\scriptsize 1 Govoni et al. (2001), 2  Bonafede et al. (2010), 3 Govoni et al. (2006), 4  Eilek \& Owen (2002),}\\
\multicolumn{6}{l}{\scriptsize 5  Feretti et al. (1999a), 6 Taylor et al. (2001), 7 Guidetti et al. (2008), * This work.}\\
\end{tabular}
\end{table*}

\section{Conclusions}

We performed Very Large Array observations at $3.6\, {\mathrm cm}$ and $6\, {\mathrm cm}$ of two radio 
galaxies located in A401 and Ophiuchus, a radio galaxy in A2142, and a radio galaxy located 
in the background of A2065.
For most of them we obtained detailed, high resolution ($\simeq$ 3\arcsec), rotation measure 
images. These $RM$ images reveal patchy structures with $RM$ fluctuations down to scales of a few kpc.
Under the assumption that the radio galaxies themselves have no effect on the measured RMs, 
these structures indicate that the intracluster magnetic fields fluctuate down to such small 
scales.

These new $RM$ data are compared with $RM$ information taken from the literature
for which similar high-resolution, extended $RM$ images are
available.

A positive trend is found between the cluster temperature and the dispersion of the $RM$ distributions 
($\sigma_{RM}$). The correlation found between the gas density 
and the gas temperature indicates that in our sample hotter clusters have higher $\sigma_{RM}$ mostly because of their higher gas density.
Moreover, although the previously known relation between 
the clusters X-ray surface brightness ($S_{X}$) at the radio galaxy location 
and $\sigma_{RM}$ is confirmed, a possible offset 
of the $\sigma_{RM}-S_{X}$ relation along the $\sigma_{RM}$ axis, depending on the cluster temperature, if present
is comparable to the intrinsic scatter of the correlation.

Therefore, in view of the current data it is not possible to establish 
a strict link between the magnetic field strength and
the gas temperature of the intracluster medium.

\begin{acknowledgements}
We are grateful to the referee for very useful comments that improved 
this paper.
We would like to thank Julia Weratschnig for helpful discussions.
This research was partially supported by ASI-INAF I/088/06/0 - 
High Energy Astrophysics and PRIN-INAF 2008.
K.~D.~acknowledges the supported by the DFG Priority Programme 1177.
The National Radio Astronomy Observatory (NRAO)
is a facility of the National Science Foundation, operated under
cooperative agreement by Associated Universities, Inc.
This research has made use of the
NASA/IPAC Extragalactic Data Base (NED) which is operated by the JPL, 
California Institute of Technology, under contract with the National 
Aeronautics and Space Administration.

\end{acknowledgements}

\appendix
\section{Rotation measure and depolarization in a foreground Faraday screen}

Under the assumption that the $RM$ spatial variations seen in the
presented work are fully explained by external patchy screens,
we investigated if these $RM$ structural information are 
consistent with the observed depolarization levels at longer wavelengths.
Indeed, the most direct consequence arising from $RM$ structures 
which fluctuate on small scales is the beam depolarization effect.

We performed this analysis through the software FARADAY 
(Murgia et al. 2004). In particular we performed 2D Montecarlo simulations 
by assuming for the rotation measure a simple power law power spectrum of the 
type:
\begin{equation}
|RM_k|^2\propto k^{-n}
\label{bpower}
\end{equation}
For each source we simulated $RM$ images by keeping fixed $n=11/3$, 
corresponding to the Kolmogorov power law index for the parent
magnetic field power spectrum, 
and by varying the minimum and maximum 
scale of the $RM$ fluctuations ($\Lambda_{min}$, $\Lambda_{max}$).
The computational grid was $1024 \times 1024$ pixel$^2$ and
the pixel-size was fixed to 1 kpc. The resulting field of view
was then $\sim$ 1024$\times$1024 kpc$^2$, that was enough to recover the
projected size of the sources and to properly sample the $RM$ power
spectrum scales. We then added to the simulated $RM$ images a Gaussian noise
having $\sigma_{noise}=Err_{fit}$, in order to mimic the effect of the
noise in the observed $RM$ images and we then added an $RM$ offset 
corresponding to the $RM$ Galactic contribution 
in the sources location. Finally, the simulated $RM$ images 
were convolved with a Gaussian function having $FWHM$ equal to the beam 
of the observations, and blanked as the observations.

We performed several 2D simulations of synthetic $RM$ images 
to evaluate the range of $\Lambda_{min}$-$\Lambda_{max}$ which could explain the observed $RM$ images.
Fig. A.1 (left panels) shows some examples of simulated $RM$ images, 
which we found to reproduce quite well the observed $RM$ distributions.

To quantify the two dimensional fluctuations of $RM$ and to compare the
simulations with the observations, we used the structure 
function as a statistical tool.
The $RM$ structure function is defined by
\begin{equation}
S(dx,dy)=\langle [RM(x,y)-RM(x+dx,y+dy)]^2\rangle_{(x,y)},
\end{equation}
where $=\langle \rangle_{(x,y)}$ indicates that the average is taken
over all the positions $(x,y)$ in the $RM$ image. Blank pixels were
not considered in the statistics. The structure function $S(r)$ was
then computed by radially averaging $S(dx,dy)$ over regions of
increasing size of radius $r=\sqrt{dx^2+dy^2}$.
In the middle panels of Figs. A.1 we compare the observed (black points) 
and the simulated (grey-dashed) $RM$ structure function.  

In the case of an external Faraday screen, a reduction
of the fractional polarization at longer wavelengths is expected 
if the minimum scale of the $RM$ 
fluctuation ($\Lambda_{min}$) extend to small scales 
(beam depolarization).
Once a good representation of the $RM$ structures have been found,
we investigated the expected beam depolarization  to check the consistency 
with the observed polarization at different wavelengths.
In particular, we produced, at each observing frequency, 
the expected $Q$ and $U$ images corresponding 
to the simulated $RM$, and we taken into account 
the beam depolarization effects by
convolving $Q$ and $U$ to the appropriate resolution. The 
predicted and observed mean degrees of polarization are then compared. 

For each source we found a combination of model parameters
$\Lambda_{min}$, $\Lambda_{max}$ that gives a reasonable 
representation of the $RM$ structure across the source 
as well as of the depolarization at longer wavelengths.
The values of $\Lambda_{min}$ and $\Lambda_{max}$
we found for these sources are consistent with those found in the
literature for a few other 
galaxy clusters (e.g. Murgia et al. 2004, Govoni et al. 2006, 
Guidetti et al. 2008, Bonafede et al. 2010). 
Therefore, even if we cannot exclude a possible presence
of an internal contribution to the observed $RM$ images, the 2D Montecarlo 
simulations indicate that both $RM$ and polarization data 
can be explained by the presence of a foreground Faraday screen.

The different magnetic field scales $\Lambda_{B}$ of the clusters 
can be compared by calculating the magnetic field autocorrelation length,  
which takes into account the minimun and maximum scale of fluctuations
($\Lambda_{min}$, $\Lambda_{max}$), and the index of 
the power spectrum ($n$).
Following the magnetic field autocorrelation length formula 
presented by En{\ss}lin \& Vogt (2003), for the
new data analyzed here we found a magnetic field 
autocorrelation length variation of a factor of 9.
This difference may influence the $\sigma_{RM}$
values of about factor of 3 ($\sigma_{RM}\propto\sqrt {\Lambda_{B}}$).
However, we note that the exact determination 
of the magnetic field power spectrum 
is beyond the scope of this paper, and the 
numbers indicated in the caption of Fig. A.1 should be considered 
as indicative.

\begin{figure*}
\vspace{21cm}
\includegraphics{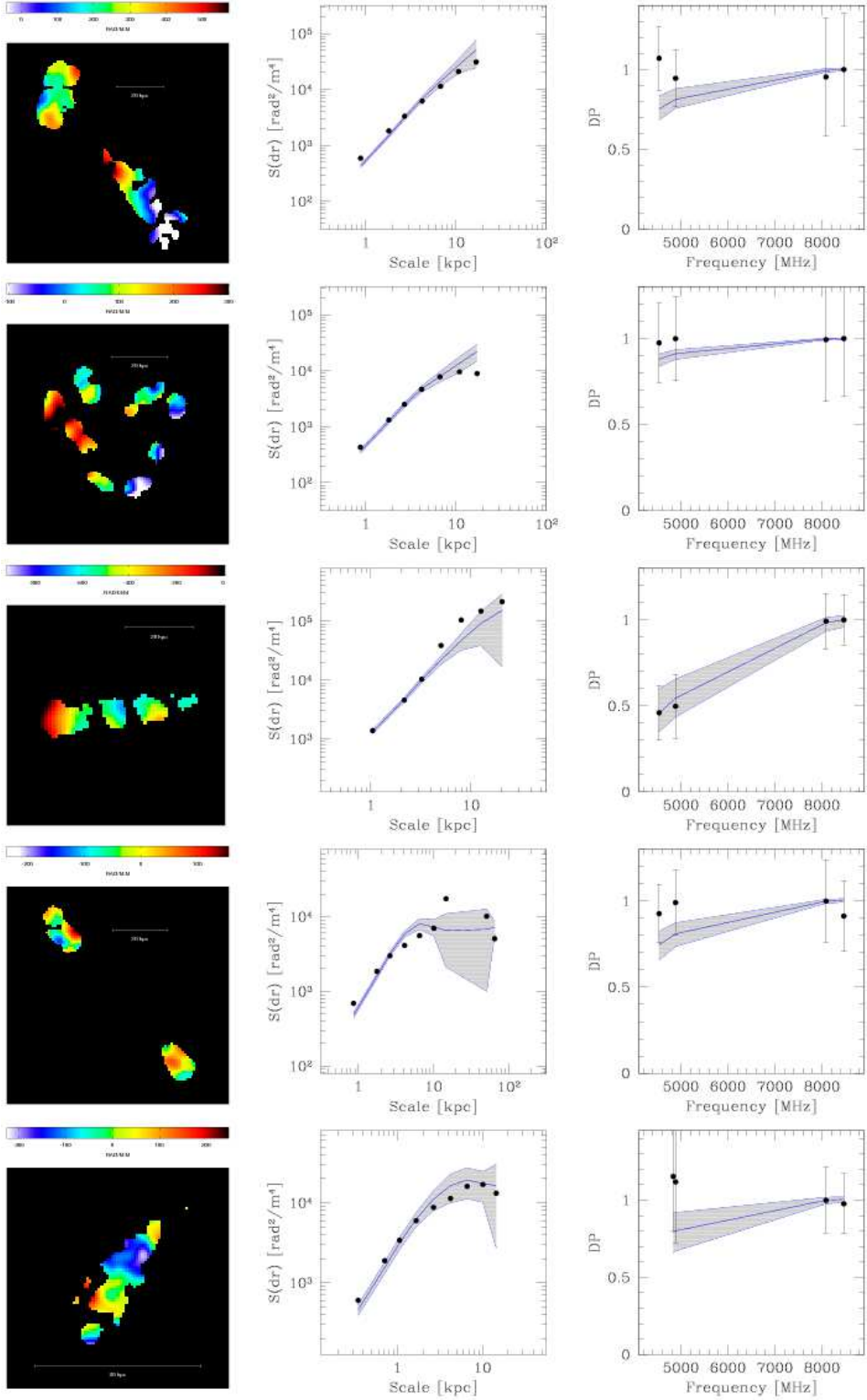}
\caption[]{
Left: Examples of simulated $RM$ images for the sources A401A, A401B, A2142, A2065, OPHIB. The simulated images have the same colour scale and resolution of the observed $RM$ images. 
They have been obtained by simulating a $RM$ power 
spectrum with a spectral index n=11/3 and a minimum scale of the $RM$ fluctuations $\Lambda_{min}$=5 kpc. 
The maximum scale of the $RM$ fluctuations is 
$\Lambda_{max}$=205 kpc for A401A and A2142,
$\Lambda_{max}$=128 kpc for A401B, 
$\Lambda_{max}$=15 kpc for OPHIB and A2065.
Middle: Structure functions of the simulated (grey) and observed (black points) $RM$ images. 
Right: Depolarization plots of the simulations (grey) and observations 
(black points). The depolarization is defined as $DP(\nu)=FPOL_{\nu}/FPOL_{8465 {\mathrm MHz}}$.
}
\label{simul}
\end{figure*}

\end{document}